\documentclass[3p]{elsarticle}

\usepackage{lineno}

\usepackage{bbm}
\usepackage{amsmath}
\usepackage{amssymb}
\usepackage{subfigure}
\usepackage{epsfig}
\usepackage{siunitx}
\DeclareSIUnit\Molar{M}     

\usepackage{color}

\usepackage{hyperref}   

\makeatletter 

\def\input@path{{01_sections/}{02_figures/}}

\newcommand{\mb}[1]{\ensuremath{\boldsymbol{#1}}}

\newcommand\DeclareBoldMathCommand[2]{%
  \protected@edef\@tempb{%
    \noexpand\DeclareRobustCommand{\csname #1\endcsname}{\boldsymbol{\ensuremath{#2}}}}
  \@tempb}
\@tfor\AM@letter:=abcdefghijklmnopqrstuvwxyz%
\do{\DeclareBoldMathCommand{v\AM@letter}{\AM@letter}}
\@tfor\AM@letter:=ABCDEFGHIJKLMNOPQRSTUVWXYZ%
\do{\DeclareBoldMathCommand{v\AM@letter}{\AM@letter}}
\DeclareBoldMathCommand{vnull}{0}
\DeclareBoldMathCommand{vone}{1}
\DeclareBoldMathCommand{valpha}{\alpha}
\DeclareBoldMathCommand{vbeta}{\beta}
\DeclareBoldMathCommand{vgamma}{\gamma}
\DeclareBoldMathCommand{vdelta}{\delta}
\DeclareBoldMathCommand{vepsilon}{\epsilon}
\DeclareBoldMathCommand{vvarepsilon}{\varepsilon}
\DeclareBoldMathCommand{vzeta}{\zeta}
\DeclareBoldMathCommand{veta}{\eta}
\DeclareBoldMathCommand{vtheta}{\theta}
\DeclareBoldMathCommand{vvartheta}{\vartheta}
\DeclareBoldMathCommand{viota}{\iota}
\DeclareBoldMathCommand{vkappa}{\kappa}
\DeclareBoldMathCommand{vlambda}{\lambda}
\DeclareBoldMathCommand{vmu}{\mu}
\DeclareBoldMathCommand{vnu}{\nu}
\DeclareBoldMathCommand{vpi}{\pi}
\DeclareBoldMathCommand{vxi}{\xi}
\DeclareBoldMathCommand{vrho}{\varrho}
\DeclareBoldMathCommand{vsigma}{\sigma}
\DeclareBoldMathCommand{vtau}{\tau}
\DeclareBoldMathCommand{vupsilon}{\upsilon}
\DeclareBoldMathCommand{vphi}{\varphi}
\DeclareBoldMathCommand{vchi}{\chi}
\DeclareBoldMathCommand{vpsi}{\psi}
\DeclareBoldMathCommand{vomega}{\omega}
\DeclareBoldMathCommand{vGamma}{\Gamma}
\DeclareBoldMathCommand{vDelta}{\Delta}
\DeclareBoldMathCommand{vTheta}{\Theta}
\DeclareBoldMathCommand{vLambda}{\Lambda}
\DeclareBoldMathCommand{vXi}{\Xi}
\DeclareBoldMathCommand{vPi}{\Pi}
\DeclareBoldMathCommand{vSigma}{\Sigma}
\DeclareBoldMathCommand{vUpsilon}{\Upsilon}
\DeclareBoldMathCommand{vPhi}{\Phi}
\DeclareBoldMathCommand{vPsi}{\Psi}
\DeclareBoldMathCommand{vOmega}{\Omega}

\newcommand\DeclareDiscreteBoldMathCommand[2]{%
  \protected@edef\@tempc{%
    \noexpand\DeclareRobustCommand{\csname #1\endcsname}{\boldsymbol{\mathrm{#2}}}}
  \@tempc}
\@tfor\AM@letter:=abcdefghijklmnopqrstuvwxyz%
\do{\DeclareDiscreteBoldMathCommand{vd\AM@letter}{\AM@letter}}
\@tfor\AM@letter:=ABCDEFGHIJKLMNOPQRSTUVWXYZ%
\do{\DeclareDiscreteBoldMathCommand{vd\AM@letter}{\AM@letter}}
\DeclareDiscreteBoldMathCommand{vdnull}{0}
\DeclareDiscreteBoldMathCommand{vdone}{1}
\DeclareDiscreteBoldMathCommand{vdalpha}{\alpha}
\DeclareDiscreteBoldMathCommand{vdbeta}{\beta}
\DeclareDiscreteBoldMathCommand{vdgamma}{\gamma}
\DeclareDiscreteBoldMathCommand{vddelta}{\delta}
\DeclareDiscreteBoldMathCommand{vdepsilon}{\epsilon}
\DeclareDiscreteBoldMathCommand{vdvarepsilon}{\varepsilon}
\DeclareDiscreteBoldMathCommand{vdzeta}{\zeta}
\DeclareDiscreteBoldMathCommand{vdeta}{\eta}
\DeclareDiscreteBoldMathCommand{vdtheta}{\theta}
\DeclareDiscreteBoldMathCommand{vdvartheta}{\vartheta}
\DeclareDiscreteBoldMathCommand{vdiota}{\iota}
\DeclareDiscreteBoldMathCommand{vdkappa}{\kappa}
\DeclareDiscreteBoldMathCommand{vdlambda}{\lambda}
\DeclareDiscreteBoldMathCommand{vdmu}{\mu}
\DeclareDiscreteBoldMathCommand{vdnu}{\nu}
\DeclareDiscreteBoldMathCommand{vdpi}{\pi}
\DeclareDiscreteBoldMathCommand{vdxi}{\xi}
\DeclareDiscreteBoldMathCommand{vdrho}{\varrho}
\DeclareDiscreteBoldMathCommand{vdsigma}{\sigma}
\DeclareDiscreteBoldMathCommand{vdtau}{\tau}
\DeclareDiscreteBoldMathCommand{vdupsilon}{\upsilon}
\DeclareDiscreteBoldMathCommand{vdphi}{\varphi}
\DeclareDiscreteBoldMathCommand{vdchi}{\chi}
\DeclareDiscreteBoldMathCommand{vdpsi}{\psi}
\DeclareDiscreteBoldMathCommand{vdomega}{\omega}
\DeclareDiscreteBoldMathCommand{vdGamma}{\Gamma}
\DeclareDiscreteBoldMathCommand{vdDelta}{\Delta}
\DeclareDiscreteBoldMathCommand{vdTheta}{\Theta}
\DeclareDiscreteBoldMathCommand{vdLambda}{\Lambda}
\DeclareDiscreteBoldMathCommand{vdXi}{\Xi}
\DeclareDiscreteBoldMathCommand{vdPi}{\Pi}
\DeclareDiscreteBoldMathCommand{vdSigma}{\Sigma}
\DeclareDiscreteBoldMathCommand{vdUpsilon}{\Upsilon}
\DeclareDiscreteBoldMathCommand{vdPhi}{\Phi}
\DeclareDiscreteBoldMathCommand{vdPsi}{\Psi}
\DeclareDiscreteBoldMathCommand{vdOmega}{\Omega}

\providecommand*{\dd}{%
  \@ifnextchar^{\@dd}{\@dd^{}}}
\def\@dd^#1{%
  \mathop{\mathrm{\mathstrut d}}%
  \nolimits^{#1}\dd@gobblespace}
\def\dd@gobblespace{%
  \futurelet\diffarg\dd@opspace}
\def\dd@opspace{%
  \let\dd@space\!%
  \ifx\diffarg(
\let\dd@space\relax%
\else%
\ifx\diffarg[
\let\dd@space\relax%
\else%
\ifx\diffarg\{%
\let\dd@space\relax%
\fi%
\fi%
\fi%
\dd@space}

\newcommand{\Frac}{%
  \@ifnextchar[
  {\Frac@i}
  {\Frac@ii}}
\newcommand{\Frac@i}{}
\def\Frac@i[#1]#2#3{%
  \genfrac{}{}{#1}{}{\displaystyle{#2}}{\displaystyle{#3}}}
\newcommand{\Frac@ii}[2]{\frac{\displaystyle{#1}}{\displaystyle{#2}}}
      \newcommand{\diff@diffspace}{\,}
\newcommand{\diff@mathfrac}[2]{\frac{#1}{#2}}
\newcommand{\diff@mathFrac}[2]{\Frac{#1}{#2}}
\newcommand{\diff@textfrac}[2]{%
  \bgroup #1\egroup\mkern-1mu/\mkern-1mu\bgroup #2\egroup}
\newcommand{\diff}{%
  \global\let\diff@diffop\dd
  \global\let\diff@frac\diff@mathfrac
  \@ifnextchar[
  {\diff@i}
  {\diff@ii}}
\newcommand{\Diff}{%
  \global\let\diff@diffop\dd
  \global\let\diff@frac\diff@mathFrac
  \@ifnextchar[
  {\diff@i}
  {\diff@ii}}
\newcommand{\tdiff}{%
  \global\let\diff@diffop\dd
  \global\let\diff@frac\diff@textfrac
  \@ifnextchar[
  {\diff@i}
  {\diff@ii}}
\newcommand{\pdiff}{%
  \global\let\diff@diffop\partial
  \global\let\diff@frac\diff@mathfrac
  \@ifnextchar[
  {\diff@i}
  {\diff@ii}}
\newcommand{\Pdiff}{%
  \global\let\diff@diffop\partial
  \global\let\diff@frac\diff@mathFrac
  \@ifnextchar[
  {\diff@i}
  {\diff@ii}}
\newcommand{\tpdiff}{%
  \global\let\diff@diffop\partial
  \global\let\diff@frac\diff@textfrac
  \@ifnextchar[
  {\diff@i}
  {\diff@ii}}

\newcommand*{\diff@i}{}
\def\diff@i[#1]#2#3{\eval{\diff@ii{#2}{#3}}_{#1}}

\newcommand*{\diff@ii}[2]{%
  \begingroup
  \toks0={}\count0=0
  \diff@degree #2\diff@degree
  \diff@frac{\diff@diffop\ifnum\count0>1^{\the\count0}\fi\diff@diffspace#1}%
  {\the\toks0}%
  \endgroup}

\newcommand*{\diff@degree}[1]{%
  \ifx #1\diff@degree \expandafter\diff@stopd
  \else \expandafter\diff@addd \fi #1^1$#1\diff@addd}
\newcommand{\diff@stopd}{}
\def\diff@stopd #1\diff@addd{}
\newcommand*{\diff@addd}{}
\def\diff@addd #1^#2#3$#4\diff@addd{%
  \advance\count0 #2
  \toks0=\expandafter{\the\toks0%
    {\diff@diffop\diff@diffspace #4}%
    \diff@diffspace}\diff@degree}

\def\rs#1{\@ifnextchar[
  {\@rs{#1}}{\@@rs{#1}}}
\def\@rs#1[#2]#3{\mathinner{%
    \setbox\@ne\hbox{$\displaystyle{\vphantom{#3}}#1{#3}\m@th$}%
    \setbox\tw@\hbox{$\displaystyle{#3}#2\m@th$}%
    \hskip\wd\@ne\hskip-\wd\tw@\mathord{\hskip\wd\tw@\hskip-\wd\@ne%
      {\vphantom{#3}}#1{#3}#2}}}
\def\@@rs#1#2{\mathinner{%
    \setbox\@ne\hbox{$\displaystyle{\vphantom{#2}}#1{#2}\m@th$}%
    \hskip\wd\@ne\mathord{\hskip-\wd\@ne%
      {\vphantom{#2}}#1{#2}}}}

\newcommand*{\norm}[1]{\mathinner{\Vert#1\Vert}}

\definecolor{notecolor}{cmyk}{0,1,1,.2}
\newcommand*\AM@notesname{Notes}

\makeatother

\modulolinenumbers[5]

\graphicspath{{02_figures/}}

\journal{International Journal of Solids and Structures}

\begin{document}

\begin{frontmatter}


\title{Generalized Section-Section Interaction Potentials in the Geometrically Exact Beam Theory: 
Modeling of Intermolecular Forces, Asymptotic Limit as Strain-Energy Function, and Formulation of Rotational Constraints}

\author[lnm]{Christoph Meier\corref{cor1}}
\ead{meier@lnm.mw.tum.de}
\author[lnm]{Maximilian J. Grill}
\author[lnm]{Wolfgang A. Wall}

\address[lnm]{Institute for Computational Mechanics, Technical University of Munich, Boltzmannstrasse 15, 85748 Garching b. M\"unchen, Germany}

\cortext[cor1]{Corresponding author}

\begin{abstract}

The present contribution proposes a universal framework to formulate generalized section-section interaction potentials (SSIP) within the geometrically exact beam theory. By exploiting the fundamental kinematic assumption of undeformable cross-sections, an objective (i.e., frame-invariant) description of SSIPs via a minimal set of six (translational and rotational) relative coordinates, either in spatial or in material form, is proposed. Based on work-pairing, work-conjugated section-section interaction forces and moments, either in spatial or in material form, are identified that can be consistently derived from a variational principle. Interestingly, it is shown that hyperelastic stored-energy functions relating the deformation measures and stress-resultants of the well-known geometrically exact Simo-Reissner beam theory can also be identified as SSIPs when considering the asymptotic limit of small relative distances and rotations between the interacting cross-sections. Moreover, the proposed variational problem formulation is demonstrated to be of a very general nature, thus allowing for the formulation of translational and rotational constraints between arbitrarily oriented cross-sections based on either a penalty or a Lagrange multiplier potential. Possible applications include fiber-based structures and materials in technical and biological systems, where the proposed approach allows to model short- or long-ranged inter-molecular (e.g., electrostatic, van der Waals or repulsive steric) interactions between fibers in geometrically complex arrangements and to formulate translational and rotational coupling constraints between different fibers (e.g., cross-linked polymer chains) or between fibers and a matrix phase (e.g., fiber-reinforced composites).
\end{abstract}


\end{frontmatter}




\section{Introduction}

There are countless fields of application, either in technical or biological systems, for fiber-based structures and materials. Typically, the behavior of such systems is governed by the mechanical properties of individual fibers, their geometrical arrangement and by local (mesoscale) interactions among fibers or between fibers and a matrix material. In fiber-based technical structures such as ropes, cables, meshes or webbings, the global structural behavior is mostly governed by the geometrical fiber arrangement and their mechanical contact interaction (see e.g.~\cite{durville2010, Kulachenko2012, Weeger2016, Meier2017b, Pattinson2019}). Fiber-based technical materials comprise composites based on carbon, glass or polymer reinforcement fibers embedded in plastic, metal or ceramic matrix materials, but also concrete structures with steel reinforcement (see e.g.~\cite{Mattheij2000, durville2007, steinbrecher2020, Khristenko2021}). The local mechanical load transfer and in particular the mechanisms of fatigue and failure in such systems is strongly related to the mechanical coupling between fibers and matrix. Eventually, there is an abundance and manifoldness of biological, fiber-like structures on the nano- and microscale, including filamentous actin, collagen, and DNA, among others. These slender, deformable fibers form a variety of complex, hierarchical assemblies such as networks (e.g.~cytoskeleton, extracellular matrix, mucus) or bundles (e.g.~muscle, tendon, ligament), which are crucial for numerous essential processes in the human body and other biological systems (see e.g.~\cite{Castro2011,Gautieri2012,Sauer2009,lindstrom2010biopolymer, muller2015resolution,Negi2018,Goodrich2018,GrillParticleMobilityHydrogels,GrillPeelingPulloff,Eichinger2021,BundlesPNAS}). At these length-scales, inter-molecular interactions, e.g., due to electrostatic, van der Waals or repulsive steric potentials, are often the key to the functionality and behavior on the system level. In the present work, a general modeling framework is proposed for such inter-molecular interactions between slender fibers based on the geometrically exact beam theory. 

There are many contributions that focus on the modeling of molecular interactions between arbitrarily shaped, solid bodies in 3D space~\cite{Argento1997,Sauer2007a,Sauer2009a,Sauer2013,Fan2015,Du2019,Mergel2019}. However, the direct evaluation of the interaction potential
between two general bodies in 3D space requires to integrate molecule densities over their volumes, generally leading to a sixfold integral (two nested 3D integrals) that has to be solved numerically. For the simulation of representative, i.e., sufficiently large, 3D systems of slender fibers more efficient reduced-order models are needed that still account for the molecular interactions in a consistent manner. While there is a large number of articles~\cite{durville2010,Kulachenko2012,Meier2017b,wriggers1997,litewka2005,Chamekh2014,GayNeto2016a,Konyukhov2016,Weeger2017,meier2016,Meier2017, bosten2022mortar} considering macroscale contact interaction between slender fibers respectively beams, comparable formulations for microscale molecular interactions are still missing to a large extent. Important steps into this direction have been made by the works~\cite{Sauer2009,Sauer2014,Schmidt2015}, however limited to the interaction of fibers respectively beams with a rigid half-space.

Based on the fundamental kinematic assumption of undeformable fiber cross-sections, as typically applied in mechanical beam theories, the authors recently proposed a modeling approach based on section-section interaction potentials (SSIP), describing the net interaction between two cross-sections of the considered fibers~\cite{GrillSSIP}. In this work, exemplary closed-form analytical solutions for the required SSIP laws could be derived for different long-ranged (e.g., electrostatic) and short-ranged (e.g., van der Waals adhesion and steric repulsion) interactions. Thereto, circular cross-section shapes, homogeneous molecule distributions as well as the asymptotic limitting cases of either large distances for long-ranged or small distances for short-ranged interactions have been considered. Due to the pre-calculated analytical representation of section-section interaction potentials, this SSIP approach only required the twofold integration along the fiber length directions to be performed numerically, thus reducing the computational effort by several orders of magnitude as compared to a direct numerical evaluation of the underlying sixfold integral. Based on an asymptotically consistent model-order reduction for the case of short-ranged potentials, the computational complexity has recently been further reduced from double to single numerical integration~\cite{Grill2022a,Grill2022b}.

The aforementioned assumptions (i.e., circular cross-sections, homogeneous molecule distributions, asymptotic limits of either large or small distances) allow for pleasantly simple SSIP laws that only require one scalar relative coordinate, i.e., the cross-section centroid distance, to fully describe the interaction kinematics, but also limit the scope of applicability. In the present work, the SSIP concept is generalized to a universal framework allowing for the interaction of cross-sections of arbitrary shape, inhomogeneous molecule distributions and arbitrary molecule-molecule interaction laws, consistently embedded into the framework of the geometrically exact beam theory~\cite{reissner1972, simo1985, simo1986, Cardona1988, ibrahimbegovic1995computational, Crisfield1999, jelenic1999, betsch2002frame, leyendecker2006objective, romero2004, romero2008, cesarek2012, Bauchau2014, sonneville2014, Meier2014, Meier2019}. 

In a first step, by exploiting the fundamental kinematic assumption of undeformable cross-sections, an objective (i.e., frame-invariant) description of SSIPs via a minimal set of six (translational and rotational) relative coordinates, either in spatial or in material form, is proposed. Based on work-pairing, work-conjugated section-section interaction forces and moments, either in spatial or in material form, are identified that can be consistently derived from a variational principle. Interestingly, it is shown that hyperelastic stored-energy functions relating the deformation measures and stress-resultants of the well-known geometrically exact Simo-Reissner beam theory can also be identified as SSIPs when considering the asymptotic limit of small relative distances and rotations between the interacting cross-sections. Moreover, the proposed variational problem formulation is demonstrated to be of a very general nature, thus allowing for the formulation of translational and rotational constraints between arbitrarily oriented cross-sections based on either a penalty or a Lagrange multiplier potential. Possible applications include fiber-based structures and materials in technical and biological systems, where the proposed approach allows to model short- or long-ranged inter-molecular (e.g., electrostatic, van der Waals or repulsive steric) interactions between arbitrarily arranged fibers and to formulate translational and rotational coupling constraints between different fibers (e.g., cross-linked polymer chains) or between fibers and a matrix phase (e.g., fiber-reinforced composites).

The remainder of this work is organized as follows: Sections~\ref{sec:rotgroup} and~\ref{geo_exact} briefly introduce the theory of large rotations and the geometrically exact beam theory. Section~\ref{sec:SSIP} introduces the concept of generalized section-section interaction potentials including the definition of \textit{generalized deformation measures}, i.e., a minimal set of translational and rotational relative coordinates between two cross-sections, and the derivation of interaction forces and moments on basis of a variational principle. Section~\ref{sec:specialSSIP} presents special cases of SSIPs with high practical relevance, in particular hyperelastic stored-energy functions of the geometrically exact Simo-Reissner beam theory identified as asymptotic limiting case of SSIPs when considering small relative distances and rotations between the interacting cross-sections as well as SSIPs in form of penalty or Lagrange multiplier potentials that can be employed to enforce general translational and rotational constraints. Eventually, the main results of this contribution are summarized in Section~\ref{sec:summary}.

\section{The rotation group SO(3)}
\label{sec:rotgroup}

In this Section, some basics from the theory of large rotations, as far as required for this work, will be recapitulated. For a more comprehensive overview, the interested reader is referred to~\cite{simo1986, Cardona1988, ibrahimbegovic1995computational, romero2004, Meier2019}. In the following, a global Cartesian frame $\ve^1, \ve^2, \ve^3\in \Re^3$ and a local orthonormal frame $\vg^{1}, \vg^{2}, \vg^{3} \in \Re^3$ are considered. The rotation from the global to the local frame is described by the rotation tensor $\vLambda \in S\!O(3)$:
\begin{align}
\label{largerotations_lambda}
  \vg^{i}=\vLambda \ve^{i}  \quad \text{with} \quad \vLambda = \vg^{j} \otimes \ve^{j} = [\vg^{1},\vg^{2},\vg^{3}]_{\ve^{j}} \quad \text{for} \quad  i,j=1,2,3.
\end{align}
Throughout this work, the summation convention over repeated indices holds. Moreover, the index near a matrix representation (e.g., the index $[.]_{\ve^{j}}$ in equation~\eqref{largerotations_lambda}) denotes the basis in which the associated tensor is represented. The rotation tensor $\vLambda$ can be identified as element of the Special Orthogonal group $S \! O(3)$:
\begin{align}
\label{largerotations_SO3}
  S \! O(3):= \{ \vLambda \in \Re^{3\!\times\!3} | \vLambda^T\!\vLambda = \vI_3, \text{det}(\vLambda)~=~1  \}.
\end{align}
In fact, $S \! O(3)$ can be identified as Lie group with associated Lie algebra $so(3)$, which is defined as:
\begin{align}
\label{largerotations_so3}
  so(3):= \{ \vS ( \va ) | \vS ( \va )=- \vS ( \va )^T \, \forall \, \va \in \Re^3  \}.
\end{align}
Thus, $so(3)$ represents the set of skew symmetric tensors with $\vS ( \va ) \vb \!=\! \va\!\times\!\vb \,\, \text{for} \,\, \va,\vb  \!\in\! \Re^3$. The isomorphism between $so(3)$ and $\Re^3$ enables a unique representation of $\vS( \va ) \in so(3)$ by the vector $\va \in \Re^3$ denoted as axial vector. Throughout this work, a parametrization of the rotation tensor $\vLambda$ based on a rotation (pseudo-) vector $\boldsymbol{\psi} \!\in\! \Re^3$ will be considered. This parametrization can be stated by the well-known Rodrigues formula:
\begin{align}
\label{largerotations_rotrigues}
  \vLambda(\boldsymbol{\psi})  = \exp{  \! ( \mb{S}(   \boldsymbol{\psi}  ) )   } = 
  \left[
  \vI  +  \sin{\psi} \vS(\boldsymbol{e_{\boldsymbol{\psi}}}) +  (1-\cos{\psi}) \vS(\boldsymbol{e_{\boldsymbol{\psi}}}) \vS(\boldsymbol{e_{\boldsymbol{\psi}} })
  \right].
\end{align}
Here, $\psi \!=\! ||\boldsymbol{\psi}||$ is the scalar rotation angle and $\ve_{\boldsymbol{\psi}} \!=\! \boldsymbol{\psi}/||\boldsymbol{\psi}||$ the axis of rotation. Moreover, the notion $\exp{  \! ( \mb{S}(   \boldsymbol{\psi}  ) )   }$ refers to the alternative representation of~\eqref{largerotations_rotrigues} (as series expansion) based on the matrix exponential map. The rotation vector of a given rotation tensor can be extracted by employing Spurrier's algorithm~\cite{spurrier1978}. To simplify notation, the abbreviation $\text{rv}(.)\!:=\!(\mb{S}^{-1} \circ  \exp^{-1})(.)$ is used for this extraction:
\begin{align}
\label{largerotations_extraction}
\begin{split}
\boldsymbol{\psi} \!=\! \text{rv}(\boldsymbol{\Lambda}) \quad \Leftrightarrow \quad
\exp{\!(\mb{S}(\boldsymbol{\psi}))} \!=\! \boldsymbol{\Lambda}.
\end{split}
\end{align}
If a subsequent rotation $ \exp{ \! ( \mb{S}(\Delta \boldsymbol{\theta}) )} $ by a finite angle $\Delta \boldsymbol{\theta} \!\in\! \Re^3$ is superimposed 
onto a given triad with rotation vector $\boldsymbol{\psi}$ according to $\mb{\Lambda}=\exp{ \! ( \mb{S}( \boldsymbol{\psi} )})$, the resulting triad $\mb{\Lambda}_n$ with rotation vector $\boldsymbol{\psi}_n$ reads
\begin{align}
\label{largerotations_leftcompound}
  \mb{\Lambda}_n = \exp{ \! ( \mb{S}( \boldsymbol{\psi}_n ) )} = \exp{ \! ( \mb{S}( \boldsymbol{\psi} \!+\! \Delta \boldsymbol{\psi} ) )} = \exp{ \! ( \mb{S}( \Delta \boldsymbol{\theta}) )}  \exp{ \! ( \mb{S}( \boldsymbol{\psi} )}),
\end{align}
where the rotation vectors associated with the successive rotations are not additive, i.e., $\boldsymbol{\psi}_n \neq \boldsymbol{\psi}+\Delta \boldsymbol{\theta}$ or $\Delta \boldsymbol{\theta} \neq \Delta \boldsymbol{\psi}$. Consequently, $\Delta \boldsymbol{\theta}$ is commonly referred to as spatial/left multiplicative rotation increment. Since $S \! O(3)$ is also non-commutative, an alternative to the left-translation update formula~\eqref{largerotations_leftcompound} can be formulated, which is 
based on right-multiplication with the material/right multiplicative rotation increment $\Delta \boldsymbol{\Theta} \neq \Delta \boldsymbol{\theta}$:
\begin{align}
\label{largerotations_rightcompound}
  \mb{\Lambda}_n = \exp{ \! ( \mb{S}( \boldsymbol{\psi}_n ) )} = \exp{ \! ( \mb{S}( \boldsymbol{\psi} \!+\! \Delta \boldsymbol{\psi} ) )}  = \exp{ \! ( \mb{S}( \boldsymbol{\psi} ))} \exp{ \! ( \mb{S}( \Delta \boldsymbol{\Theta}) )}.
\end{align}
With auxiliary equation~\eqref{A_Lambda1}, the spatial and material multiplicative rotation increments can be related:
\begin{align}
\label{largerotations_trafoleftandright}
  \Delta \boldsymbol{\theta} = \mb{\Lambda} \Delta \boldsymbol{\Theta}.
\end{align}
Replacing the finite rotation $\Delta \boldsymbol{\theta}$ by the infinitesimal quantity $\delta \boldsymbol{\theta}$, denoted as spatial multiplicative rotation variation or spatial spin vector, yields the following expression for the variation of the rotation tensor $\mb{\Lambda}$:
\begin{align}
\label{largerotations_deltalambdaspatial}
  \delta \mb{\Lambda} := \frac{d}{d \epsilon} \Big|_{\epsilon=0}\exp{(\epsilon \mb{S}(\delta \boldsymbol{\theta}))} \mb{\Lambda} = \mb{S}(\delta \boldsymbol{\theta}) \mb{\Lambda} \quad \text{or} \quad \delta \mb{g}^i = \delta \boldsymbol{\theta} \!\times\! \mb{g}^i.
\end{align}
Similar to~\eqref{largerotations_rightcompound}, the variation $\delta \mb{\Lambda}$ can alternatively be expressed by the material spin vector $\delta \boldsymbol{\Theta}=\mb{\Lambda}^T \delta \boldsymbol{\theta}$:
\begin{align}
\label{largerotations_deltalambdamaterial}
  \delta \mb{\Lambda} := \frac{d}{d \epsilon} \Big|_{\epsilon=0} \mb{\Lambda} \exp{(\epsilon \delta \boldsymbol{\Theta})} =\mb{\Lambda}  \mb{S}(\delta \boldsymbol{\Theta}).
\end{align}
The variation $\delta \mb{\Lambda}$ can also be expressed by an (infinitesimal) additive variation $\delta \boldsymbol{\psi}$ of the rotation vector:
\begin{equation}
 \delta \mb{\Lambda} = 
 \frac{d}{d \epsilon} \Big|_{\epsilon=0}  \exp{(\boldsymbol{\psi}+\epsilon \delta \boldsymbol{\psi})} =
 \frac{\partial \vLambda}{\partial \boldsymbol{\psi}} \delta \boldsymbol{\psi}.
\end{equation}
A relation between the additive and spatial multiplicative variation is given by the tangent operator $\mb{T}$:
\begin{align}
\label{largerotations_tmatrix}
  \delta \boldsymbol{\psi}=:\mb{T}\delta \boldsymbol{\theta}, \quad
  \mb{T}=\frac{1}{\psi^2} \mb{S}( \boldsymbol{\psi} ) \mb{S}( \boldsymbol{\psi} )^T
  + \frac{\psi/2}{\tan{(\psi/2)}} \left( \mb{I} - \frac{1}{\psi^2} \mb{S}( \boldsymbol{\psi} ) \mb{S}( \boldsymbol{\psi} )^T\right) -\frac{1}{2}\mb{S}( \boldsymbol{\psi} ).
\end{align}
With~\eqref{A_T2}, a relation between the additive and material multiplicative variation can be derived from~\eqref{largerotations_tmatrix}:
\begin{align}
\label{largerotations_tmatrix2}
  \delta \boldsymbol{\psi}=\mb{T}^T \delta \boldsymbol{\Theta}.
\end{align}
By expressing the variation of an arbitrary function $\vf(\vpsi)$ via the multiplicative rotation vector variation $\delta \vtheta$ and making use of~\eqref{largerotations_tmatrix}, we define the \textit{multiplicative derivative} $\partial_m \vf / \partial_m \vtheta$ of this function as:
\begin{align}
\label{multiplicative_derivative}
\delta \vf = \frac{\partial \vf}{\partial \vpsi} \delta \vpsi = \frac{\partial \vf}{\partial \vpsi} \vT( \vpsi) \delta \vtheta=:\frac{\partial_m \vf}{\partial_m \vtheta} \delta \vtheta 
\quad \Leftrightarrow \quad \frac{\partial_m \vf}{\partial_m \vtheta}:=\frac{\partial \vf}{\partial \vpsi} \vT( \vpsi),
\end{align}
where $\partial \vf / \partial \vpsi$ is the standard partial (or additive) derivative of $\vf(\vpsi)$. The definition of this \textit{multiplicative derivative} $\partial_m \vf / \partial_m \vtheta$ will be useful for subsequent derivations.

\section{Geometrically exact beam theory}
\label{geo_exact}

In this Section, some basics of the geometrically exact beam theory, as far as required in this work, will be recapitulated. For a more comprehensive overview, the interested reader is again referred to~\cite{simo1986, Cardona1988, ibrahimbegovic1995computational, romero2004, Meier2019}. To simplify the presentation, the following contents are limited to the static case.

\subsection{Kinematics}

In the initial (unstressed) configuration, the centerline of a beam is described by the curve $s \rightarrow \mb{r}^0(s) \in \Re^3$. Here, $s \in [0,l] \subset \Re$ is an arc-length parametrization of this curve, i.e.,  $||d \mb{r}^0(s)/ds||=1$, and $l \in \Re$ the initial length of the beam. The initial configuration is completed by a field of right-handed orthonormal triads according to $s \rightarrow \mb{g}^{10}(s), \mb{g}^{20}(s), \mb{g}^{30}(s) \in \Re^3$, which are attached to the beam cross-sections and whose orientation is defined by the rotation tensor $s \rightarrow \mb{\Lambda}^0(s) \in S\!O(3)$ according to $\mb{g}^{i0}(s)=  \mb{\Lambda}^0(s) \mb{e}^{i}$. Correspondingly, the deformed configuration of the beam is defined by the centerline curve $s \rightarrow \mb{r}(s) \in \Re^3$ and the triad field $s \rightarrow \mb{g}^{1}(s), \mb{g}^{2}(s), \mb{g}^{3}(s) \in \Re^3$ with $\mb{g}^{i}(s)=  \mb{\Lambda}(s) \mb{e}^{i}$. While the first base vector of the initial triad is aligned tangentially to the centerline curve, i.e., $\mb{g}^{10}(s)= d \mb{r}^0(s)/ds$, this does in general not apply for the deformed configuration, i.e., $\mb{g}^{1}(s) \neq d \mb{r}(s)/ds$. According to the last section, $\mb{\Lambda}(s)$ can be represented by three rotation parameters (e.g., by a rotation vector $\boldsymbol{\psi}(s)$), leading to pointwise six, three translational and three rotational, degrees of freedom. The basic kinematic assumption of undeformable cross-sections, as underlying the geometrically exact Simo-Reissner theory, states that the initial and current positions $\vX$ and $\vx$ of an arbitrary material point within the cross-section can be described as follows:
\begin{align}
\label{simoreissner_x0}
 \vX (s,\xi^2,\xi^3)=\mb{r}^{0}(s) + \underbrace{\xi^2 \mb{g}^{20}(s) + \xi^3 \mb{g}^{30}(s)}_{\mb{\xi}^0}, \quad \vx (s,\xi^2,\xi^3)=\mb{r}(s) + \underbrace{\xi^2 \mb{g}^{2}(s) + \xi^3 \mb{g}^{3}(s)}_{\mb{\xi}},
\end{align}
where $\xi^2$ and $\xi^3$ represent convective coordinates spanning the cross-section plane. Moreover, $\mb{\xi}$ is the distance between an arbitrary point within the cross-section and its centroid. In analogy to~\eqref{largerotations_deltalambdaspatial} and~\eqref{largerotations_deltalambdamaterial}, the spatial and material curvature vectors $\vk$ and $\vK$ are defined according to
\begin{align}
\label{simoreissner_spatialcurvature}
  \mb{\Lambda}^{\prime}(s) =  \mb{S}(
\mb{k}(s))\mb{\Lambda}(s) \quad \text{or} \quad  
  \mb{\Lambda}^{\prime}(s) = \mb{\Lambda}(s)\mb{S}(\mb{K}(s)) \quad \text{with} \quad
  \vk= \vLambda \vK,
\end{align}
where $(.){\prime}=d (.)/d s$ represents the arc-length derivative. The kinematics are completed by the material deformation measures $\vGamma = \vLambda^T \vr^{\prime}  \!-\! \ve^1$, representing axial tension and shear, and $\vOmega=\vK - \vK^0$, representing bending and torsion, as well as their spatial counterparts $\vgamma= \vLambda \vGamma$ and $\vomega=\vLambda \vOmega$. 

\subsection{Strong form of equilibrium}

With $\mb{\tilde{f}}$ and $\mb{\tilde{m}}$ denoting distributed external forces and moments per unit length that are acting along the beam centerline, the strong form of the static equilibrium of forces and moments reads~\cite{reissner1972, simo1985,antmann1995}:
\begin{align}
\label{simoreissner_equilibriumspatial}
\begin{split}
\mb{f}^{\prime} + \mb{\tilde{f}} & = \mb{0}, \\
\mb{m}^{\prime} + \mb{r}^{\prime} \times \mb{f} + \mb{\tilde{m}}& = \mb{0}.
\end{split}
\end{align}
In~\eqref{simoreissner_equilibriumspatial}, $\mb{f}$ and $\mb{m}$ are the force and moment stress resultants acting on the beam cross-section with area $A$. Alternatively, also a material form of the 1D equilibrium equations can be derived by inserting the material stress resultants $\mb{F}\!\!:=\!\!\mb{\Lambda}^T \!\mb{f}$ and $\mb{M}\!\!:=\!\!\mb{\Lambda}^T\! \mb{m}$ into the balance equations~\eqref{simoreissner_equilibriumspatial}.

\subsection{Variational problem statement and weak form of equilibrium}

Assuming the simplest case of hyperelastic material behavior, a length-specific stored-energy function can be postulated either as function of the material deformation measures $\bar{\pi}_{int}(\mb{\Gamma},\mb{\Omega})$ or as function of the spatial deformation measures $\tilde{\pi}_{int}(\vgamma,\vomega)$. Exemplary stored-energy functions of this type are given by:
\begin{align}
\label{storedenergyfunction}
\bar{\pi}_{int}(\mb{\Gamma},\mb{\Omega}) \!=\! \frac{1}{2}  \mb{\Gamma}^T \mb{C}_F \mb{\Gamma}  \!+\! \frac{1}{2} \mb{\Omega}^T \! \mb{C}_M \mb{\Omega}
\quad \text{or} \quad 
\tilde{\pi}_{int}(\vomega,\vgamma) \!=\! \frac{1}{2} \vgamma^T \mb{c}_f \vgamma  \!+\! \frac{1}{2} \vomega^T \! \mb{c}_m \vomega,
\end{align}
where the constant material and spatial constitutive tensors are related through $\mb{c}_f=  \vLambda \mb{C}_F  \vLambda^T$  as well as $\mb{c}_m=  \vLambda \mb{C}_M  \vLambda^T$. Let us assume that the total potential energy of a beam problem can be formulated as
\begin{align}
\label{potential_energy}
\Pi_{tot}=\Pi_{int}+\Pi_{ext}+\Pi_{mol} \quad \text{with} \quad
\Pi_{int}= \int_l \bar{\pi}_{int}(\mb{\Gamma},\mb{\Omega}) d s = \int_l \tilde{\pi}_{int}(\vgamma,\vomega)  d s,
\end{align}
where $\Pi_{ext}$ denotes the potential of external forces and moments, $\Pi_{int}$ the potential of internal forces and moments and $\Pi_{mol}$ represents molecular interaction potentials as considered in the next section. To derive the weak form of the balance equations, e.g., as basis for a subsequent finite element discretization, the variation of these energy contributions is required. Variation of the internal energy contributions, either based on the material or spatial representation of the stored-energy function in~\eqref{storedenergyfunction}, yields:
\begin{align}
\label{storedenergyfunction_variation}
\delta \Pi_{int} = \int_l \left(\vF \delta \mb{\Gamma} + \vM \delta \mb{\Omega} \right) d s = 
\int_l \left(\vf \,  \delta_o \vgamma + \vm \, \delta_o \vomega \right) d s.
\end{align}
Here, the following definitions of material and spatial force and moment stress resultants have been employed:
\begin{align}
\label{storedenergyfunction_stressresultants}
\vF= \frac{\partial \bar{\pi}_{int}(\mb{\Gamma},\mb{\Omega})}{ \partial \mb{\Gamma}}, \quad
\vM = \frac{\partial \bar{\pi}_{int}(\mb{\Gamma},\mb{\Omega})}{ \partial \mb{\Omega}}, \quad
\vf = \frac{\partial \tilde{\pi}_{int}(\vgamma,\vomega)}{ \partial \vgamma}, \quad
\vm = \frac{\partial \tilde{\pi}_{int}(\vgamma,\vomega)}{ \partial \vomega}.
\end{align}
Moreover, the variations of the material deformation measures as well as the objective variations of the spatial deformation measures as occurring in~\eqref{storedenergyfunction_variation} are defined according to:
\begin{align}
\label{storedenergyfunction_objectivevariations}
\delta \boldsymbol{\Gamma} = \mb{\Lambda}^T \! \left(\delta \mb{r}^{\prime} + \mb{r}^{\prime} \times \delta \boldsymbol{\theta} \right), \quad
\delta \boldsymbol{\Omega} = \mb{\Lambda}^T \delta \boldsymbol{\theta}^{\prime}, \quad
\delta_{o} \boldsymbol{\gamma} = \delta \mb{r}^{\prime} - \delta \boldsymbol{\theta} \times
\mb{r}^{\prime}, \quad
\delta_{o} \boldsymbol{\omega} =  \delta \boldsymbol{\theta}^{\prime}.
\end{align}
In general, the objective variation $\delta_{o}(.)$ of an arbitrary vector $\mb{a} \! \in \! \Re^3$ is defined as $\delta_{o} \mb{a}\!:=\!\delta \mb{a} \!-\! \delta \boldsymbol{\theta} \!\times\! \mb{a}$ (see e.g., \cite{simo1985}). Alternatively, the spatial weak form of the balance equations can be derived via the principle of virtual work, i.e., by multiplication of the strong form of force and moment balance~\eqref{simoreissner_equilibriumspatial} with the virtual displacements $\delta \mb{r}$ and virtual (multiplicative) rotations $\delta \boldsymbol{\theta}$ and a subsequent (two-fold) integration by parts.

\section{Generalized section-section interaction potentials}
\label{sec:SSIP}
\subsection{Kinematics}
\label{sec:kinematics}

In a next step, the molecular interaction potentials $\Pi_{mol}$ in~\eqref{potential_energy} will be further specified. To shorten notation, the subscript $(.)_{mol}$ will be omitted throughout this section. In the following, two interacting beams (initial lengths $l_1, l_2$, cross-section areas $A_1,A_2$) will be considered. Moreover, general interaction potentials $\Phi(x)$, denoted as molecule-to-molecule interaction potentials, are assumed, which describe the interaction of ensembles of molecules contained in the infinitesimal volumes $dV_1=\dd A_1 \dd s_1$ and $dV_2=\dd A_1 \dd s_1$ in a homogenized continuum sense. The distance between the volumes $dV_1$ and $dV_2$ is denoted as $x$. The overall interaction potential between the two considered beams is given by the integral of the molecule-to-molecule interaction potential $\Phi(x)$ over the total volumes $V_1$ and $V_2$ of the interacting beams:
\begin{align}
\label{potential_1}
  \Pi = \int_{l_1} \int_{l_2} \underbrace{\int_{A_1} \int_{A_2} \rho_1(\mb{\xi}_1) \rho_2(\mb{\xi}_2) \Phi(x) \dd A_2 \dd A_1}_{\pi} \dd s_2 \dd s_1.
\end{align}
Here, $\rho_i$ represents the molecule density of beam $i$. From a physical point of view, the integral of $\Phi(x)$ over the two beam cross-sections (weighted by the molecule densities $\rho_1(\mb{\xi}_1)$ and $\rho_2(\mb{\xi}_2)$) represents the net interaction between two cross-sections, and shall be denoted as section-section interaction potential $\pi$. Moreover, the distance $x$ between the interacting infinitesimal volume elements $dV_1$ and $dV_2$ is defined as:
\begin{align}
\label{x21}
  x = \norm{\vx_{21}} \quad \text{with} \quad \vx_{21} = \vx_2 - \vx_1.
\end{align}
In a next step, let us define the relative distance $\vr_{21}$ between the two cross-section centroids as well as the relative rotation $\vLambda_{21} = \exp{\!(\vS(\vpsi_{21}))}$ between the two cross-section triads:
\begin{subequations}
\label{d_and_psi}
\begin{align}
  \vr_{21} := \vr_2 - \vr_1 \quad & \Leftrightarrow \quad \vr_2 = \vr_1 + \vr_{21}\\
  \vLambda_{21} := \vLambda_2 \vLambda_1^\text{T} \quad & \Leftrightarrow \quad \vLambda_2 = \vLambda_{21} \vLambda_1 \quad \text{or} \quad \vg_2^i = \vLambda_{21} \, \vg_1^i \quad \text{for} \quad  i=1,2,3.
\end{align}
\end{subequations}
An exemplary coordinate representation w.r.t. the basis $\vLambda_1$ shall be given for these kinematic quantities:
\begin{align}
\label{d_and_psi_2}
  \vr_{21} =: r_{21}^i\vg_1^i, \quad  \vpsi_{21} =: \psi_{21}^i\vg_1, \quad  \vLambda_{21} =: \Lambda_{21}^{ij} \vg_1^{i} \otimes \vg_1^{jT}.
\end{align}
Making use of~\eqref{x21},~\eqref{d_and_psi} and~\eqref{simoreissner_x0}, the distance vector $\vx_{21}$ between two interacting volumes can be rewritten:
\begin{align}
\begin{split}
\label{x21_final}
  \vx_{21} &= \vr_2 - \vr_1 + \left( \xi_2^2 \vg_2^2 + \xi_2^3 \vg_2^3 \right) - \left( \xi_1^2 \vg_1^2 + \xi_1^3 \vg_1^3 \right)\\
  &= \vr_{21} + \vLambda_{21} \left( \xi_2^2 \vg_1^2 + \xi_2^3 \vg_1^3 \right) - \left( \xi_1^2 \vg_1^2 + \xi_1^3 \vg_1^3 \right).
\end{split}
\end{align}
When considering~\eqref{x21_final} together with~\eqref{d_and_psi_2} the following statement can be made: The distance vector $\vx_{21}$ between to given cross-section points $(\xi_1^2,\xi_1^3)$ and $(\xi_2^2,\xi_2^3)$, e.g., expressed in the basis $\vLambda_1$ (with known base vectors $g_1^i$) according to $\vx_{21}=:x_{21}^ig_1^i$, is uniquely defined by the relative position and rotation vectors $\vr_{21}$ and $\vpsi_{21}$ (used for parameterization of $\vLambda_{21}$). Consequently, as expected, after integrating over the four cross-section coordinates $\xi_i^2$ and $\xi_i^3$ in~\eqref{potential_1}, the resulting section-section interaction potential is a pure function of the six relative degrees of freedom between the interacting cross-sections, i.e., $\pi=\tilde{\pi}(\vr_{21},\vpsi_{21})$. In other words, $\vr_{21}$ and $\vpsi_{21}$ can be interpreted as a minimal set of (translational and rotational) relative coordinates, in the following also denoted as \textit{generalized spatial deformation measures}, that allow to uniquely describe the considered interaction potential. By defining the rotation tensors $\vLambda_1$ and  $\vLambda_1^T$ as push-forward and pull-back operator, also a material representation of these generalized deformation measures can be stated:
\begin{align}
\label{D_and_Psi}
  \vr_{21}:=\vLambda_1\vR_{\mathbbm{2}\mathbbm{1}}, \,\, \vpsi_{21}=\vLambda_1\vPsi_{\mathbbm{2}\mathbbm{1}}  \quad \Leftrightarrow \quad \vR_{\mathbbm{2}\mathbbm{1}}:=\vLambda_1^T\vr_{21}, \,\,\vPsi_{\mathbbm{2}\mathbbm{1}}=\vLambda_1^T\vpsi_{21}.
\end{align}
Making use of~\eqref{d_and_psi}, the rotation tensor $\vLambda_{\mathbbm{2}\mathbbm{1}} := \exp{\!(\vS(\vPsi_{\mathbbm{2}\mathbbm{1}}))}$ associated with $\vPsi_{\mathbbm{2}\mathbbm{1}}$ can be written as
\begin{align}
\label{Lambda_Psi}
  \vLambda_{\mathbbm{2}\mathbbm{1}} := \exp{\!(\vS(\vLambda_1^T \vpsi_{21}))}  = \vLambda_1^T  \exp{\!(\vS(\vpsi_{21}))} \vLambda_1 = \vLambda_1^T \vLambda_{21} \vLambda_1 = \vLambda_1^T \vLambda_2 \quad \Leftrightarrow \quad \vLambda_2 =  \vLambda_1\vLambda_{\mathbbm{2}\mathbbm{1}},
\end{align}
where~\eqref{A_Lambda1} has been exploited from the second to the third expression. Thus, the spatial quantity $\vpsi_{21}$ defines the relative rotation via left-multiplication of $\vLambda_1$, while the material quantity $\vPsi_{\mathbbm{2}\mathbbm{1}}$ defines it via right-multiplication of $\vLambda_1$. Considering the coordinate representations~\eqref{d_and_psi_2}, we get
\begin{align}
\label{D_and_Psi_2}
  \vR_{\mathbbm{2}\mathbbm{1}} = r_{21}^i\ve_1^i, \quad  \vPsi_{\mathbbm{2}\mathbbm{1}} = \psi_{21}^i\ve_1, \quad  \vLambda_{\mathbbm{2}\mathbbm{1}} = \Lambda_{21}^{ij} \ve^{i} \otimes \ve^{jT}.
\end{align}
Thus, the coordinates of the material deformation measures when expressed in the inertial frame $\ve^{i}$ are identical to the coordinates of the spatial deformation measures when expressed in the moving frame $\vg_1^{i}$. With these results at hand, also the distance vector $\vx_{21}$ 
can be back-rotated according to:
\begin{align}
\begin{split}
\label{X21_final}
  \vX_{21}:= \vLambda_1^T \vx_{21} &= \vLambda_1^T \vr_{21} + \vLambda_1^T \vLambda_{21} \left( \xi_2^2 \vg_1^2 + \xi_2^3 \vg_1^3 \right) - \vLambda_1^T \left( \xi_1^2 \vg_1^2 + \xi_1^3 \vg_1^3 \right)\\
  &=\vLambda_1^T \vr_{21} + \vLambda_1^T \vLambda_{21}  \vLambda_1 \vLambda_1^T \left( \xi_2^2 \vg_1^2 + \xi_2^3 \vg_1^3 \right) - \vLambda_1^T \left( \xi_1^2 \vg_1^2 + \xi_1^3 \vg_1^3 \right)\\
  &=\vR_{\mathbbm{2}\mathbbm{1}} +  \vLambda_{\mathbbm{2}\mathbbm{1}} \left( \xi_2^2 \ve^2 + \xi_2^3 \ve^3 \right) -  \left( \xi_1^2 \ve^2 + \xi_1^3 \ve^3 \right).
\end{split}
\end{align}
Together with~\eqref{D_and_Psi_2} it becomes clear that the coordinates of the distance vector $\vX_{21}$ when expressed in the basis $\ve_1^i$, i.e., $\vX_{21}=\vLambda_1^T \vx_{21}=:x_{21}^i \ve_1^i$, are a pure function of $\vR_{\mathbbm{2}\mathbbm{1}}$ and $\vPsi_{\mathbbm{2}\mathbbm{1}}$ (used for parameterization of $\vLambda_{\mathbbm{2}\mathbbm{1}}$). Since the back-rotation $\vX_{21}:= \vLambda_1^T \vx_{21}$ preserves the $L_2-$norm, i.e,. $\norm{\vX_{21}}=\norm{\vLambda_1^T \vx_{21}}=\norm{\vx_{21}}=x$, the molecule-to-molecule interaction potential $\Phi$ can be equivalently written as $\Phi(x)=\Phi(\norm{\vx_{21}})=\Phi(\norm{\vX_{21}})$. The section-section interaction potential $\pi$, however, will in general be a function of the vectors $(\vr_{21},\vpsi_{21})$ or  $(\vR_{\mathbbm{2}\mathbbm{1}},\vPsi_{\mathbbm{2}\mathbbm{1}})$, but not only of their norms. Thus, the corresponding functional expressions $\tilde{\pi}$ and $\bar{\pi}$ in spatial and material configuration will be different in general:
\begin{align}
\label{barbar_pi}
\tilde{\pi}(\vr_{21},\vpsi_{21})=\tilde{\pi}(\vLambda_1\vR_{\mathbbm{2}\mathbbm{1}},\vLambda_1\vPsi_{\mathbbm{2}\mathbbm{1}})={\bar{\pi}}(\vR_{\mathbbm{2}\mathbbm{1}},\vPsi_{\mathbbm{2}\mathbbm{1}}).
\end{align}
In preparation to the following variational formulation of the considered beam-to-beam interaction problem, the variation of the introduced kinematic quantities will be required. 
As demonstrated in Section~\ref{geo_exact}, these variations need to be expressed via the variations of the primary degrees of freedom of the geometrically exact beam theory, i.e., the spatial variations of the cross-section centroid positions $\delta \vr_1,\delta \vr_2$ and the spatial spin vectors $\delta \vtheta_1,\delta \vtheta_2$ associated with the cross-section triads of the beams. As shown in Appendix~\ref{appendix:variations}, the variation of the material deformation measures $\vR_{\mathbbm{2}\mathbbm{1}}$ and $\vPsi_{\mathbbm{2}\mathbbm{1}}$ is given by:
\begin{subequations}
\label{delta_material}
\begin{align}
  \delta \vR_{\mathbbm{2}\mathbbm{1}} &= \vLambda_1^T \left( \delta \vr_2 - \delta \vr_1 - \delta \vtheta_1 \times \left( \vr_2 - \vr_1 \right)  \right)\\
  \delta \vPsi_{\mathbbm{2}\mathbbm{1}} & = \vT(\vPsi_{\mathbbm{2}\mathbbm{1}}) \underbrace{\vLambda_1^T \left( \delta \vtheta_2 - \delta \vtheta_1 \right)}_{=:\delta \vtheta_{\mathbbm{2}\mathbbm{1}}},
\end{align}
\end{subequations}
where the multiplicative variation $\delta \vtheta_{\mathbbm{2}\mathbbm{1}}:=\vLambda_1^T \left( \delta \vtheta_2 - \delta \vtheta_1 \right)$ has been defined in accordance to~\eqref{largerotations_deltalambdaspatial}, i.e., it describes the variation of $\vLambda_{\mathbbm{2}\mathbbm{1}}(\vPsi_{\mathbbm{2}\mathbbm{1}})$ via left-multiplication according to $\delta \vLambda_{\mathbbm{2}\mathbbm{1}} =: \vS(\delta \vtheta_{\mathbbm{2}\mathbbm{1}}) \vLambda_{\mathbbm{2}\mathbbm{1}}$.

According to the definitions above, spatial vectors $\va$ are defined as push-forward of material vectors $\vA$ according to $\va=\vLambda_1 \vA$, where $\vLambda_1$ is the push-forward and $\vLambda_1^T$ the pull-back operator, i.e., $\vA=\vLambda_1^T \va$. Expressing a spatial vector in the moving frame $\vLambda_1$ yields the following two coordinate representations $\va=a^i \vg_1^i$ and $\vA=a^i \ve_1^i$. Now, the objective variation $\delta_o \va$ of a spatial vector shall be defined as the total variation $\delta \va$ minus the contribution from the rotation of the coordinate frame. From the total variation
\begin{align}
\label{obj_var_general}
\delta \va  = \delta (\vLambda_1 \vA) = \vLambda_1 \delta \vA + \delta \vLambda_1 \vA  = \vLambda_1 \delta \vA +\vS(\delta \vtheta_1) \vLambda_1 \vA = 
\underbrace{\vLambda_1 \delta \vA}_{=\delta a^i \vg_1^i} + \underbrace{\delta \vtheta_1 \times  \va}_{ a^i \delta \vg_1^i}
\end{align}
 the following two equivalent expressions for the objective variation of a spatial vector can be identified:
\begin{align}
\label{obj_var_general}
\delta_o \va:=\delta \va - \delta \vtheta_1 \times \va=\vLambda_1 \delta \vA.
\end{align}
With the second expression in~\eqref{obj_var_general}, the objective variations of $\vr_{21}$ and $\vpsi_{21}$ can be directly derived from~\eqref{delta_material}:
\begin{subequations}
\label{delta_o}
\begin{align}
  \delta_o \vr_{21} &:= \vLambda_1 \delta \vR_{\mathbbm{2}\mathbbm{1}} = \left( \delta \vr_2 - \delta \vr_1 - \delta \vtheta_1 \times \left( \vr_2 - \vr_1 \right)  \right) \label{delta_o_1} \\
  \delta_o \vpsi_{21} &:= \vLambda_1 \delta \vPsi_{\mathbbm{2}\mathbbm{1}}  =  \vLambda_1 \vT(\vPsi_{\mathbbm{2}\mathbbm{1}}) \vLambda_1^T \left( \delta \vtheta_2 - \delta \vtheta_1 \right)  =  \vT(\vpsi_{21}) \left( \delta \vtheta_2 - \delta \vtheta_1 \right).\label{delta_o_2}
\end{align}
\end{subequations}
In the last reformulation step, use has been made of utility equation~\eqref{A_T1}. Using the last expression in the second line,~\eqref{delta_o_2} can be reformulated to
\begin{align}
  \delta_o \vpsi_{21} =  \vT(\vpsi_{21}) \left( \delta \vtheta_2 - \delta \vtheta_1 \right) =: \vT(\vpsi_{21}) \delta_o \vtheta_{21},
\end{align}
where the objective multiplicative variation $\delta_o \vtheta_{21}:=\delta \vtheta_2 - \delta \vtheta_1$ has been defined in accordance to~\eqref{largerotations_deltalambdaspatial}:
\begin{align}
\label{multi_var_21_2}
\delta_o \vLambda_{21} =: \vS(\delta_o \vtheta_{21}) \vLambda_{21}=\frac{d}{d \epsilon} \Big|_{\epsilon=0}  \exp{(\vpsi_{21}+\epsilon \delta_o \vpsi_{21})}.
\end{align}

\begin{center}
\begin{minipage}{0.9\textwidth}
{
\noindent \textbf{Remark:} With the coordinate representations~\eqref{d_and_psi_2} and~\eqref{D_and_Psi_2}, the objective variation of the spatial relative rotation tensor $\vLambda_{21}$ can alternatively be expressed as push-forward of the variation of the material relative rotation tensor $\vLambda_{\mathbbm{2}\mathbbm{1}}$, i.e., $\delta_o \vLambda_{21}= \delta \Lambda_{21}^{ij} \vg_1^{i} \otimes \vg_1^{jT} = \vLambda_{1} \delta \vLambda_{\mathbbm{2}\mathbbm{1}} \vLambda_{1}^T$. 
Using the relation $\vLambda_{1} \delta \vLambda_{\mathbbm{2}\mathbbm{1}} \vLambda_{1}^T= 
\vLambda_{1} \frac{d}{d \epsilon} \big|_{\epsilon=0}  \exp{(\vPsi_{\mathbbm{2}\mathbbm{1}}+\epsilon \delta \vPsi_{\mathbbm{2}\mathbbm{1}})} \vLambda_{1}^T =
 \frac{d}{d \epsilon} \big|_{\epsilon=0}  \exp{(\vLambda_{1} \vPsi_{\mathbbm{2}\mathbbm{1}}+\epsilon \vLambda_{1} \delta \vPsi_{\mathbbm{2}\mathbbm{1}})} =
 \frac{d}{d \epsilon} \big|_{\epsilon=0}  \exp{(\vpsi_{21}+\epsilon \delta_o \vpsi_{21})} $, the objective variation of the spatial relative rotation tensor can be expressed as 
 $\delta_o \vLambda_{21}=\frac{d}{d \epsilon} \big|_{\epsilon=0}  \exp{(\vpsi_{21}+\epsilon \delta_o \vpsi_{21})}$, i.e., the objective additive variation of the relative rotation vector $\delta_o \vpsi_{21}$ is the additive rotation vector increment associated with the objective variation of the spatial relative rotation tensor. Finally, using the relations $\delta_o \vLambda_{21}=\vLambda_{1} \delta \vLambda_{\mathbbm{2}\mathbbm{1}} \vLambda_{1}^T=
 \vLambda_{1} \vS(\delta \vtheta_{\mathbbm{2}\mathbbm{1}}) \vLambda_{\mathbbm{2}\mathbbm{1}} \vLambda_{1}^T=
 \vLambda_{1} \vS(\delta \vtheta_{\mathbbm{2}\mathbbm{1}}) \vLambda_{1}^T \vLambda_{1} \vLambda_{\mathbbm{2}\mathbbm{1}} \vLambda_{1}^T=
 \vS(\delta_o \vtheta_{21})  \vLambda_{21}$, the objective variation of the spatial relative rotation tensor can alternatively be expressed as 
 $\delta_o \vLambda_{21}=\vS(\delta_o \vtheta_{21})  \vLambda_{21}$, i.e., the objective multiplicative variation $\delta_o \vtheta_{21}$ is the multiplicative rotation vector increment associated with the objective variation of the spatial relative rotation tensor.
}
\end{minipage}
\end{center}

\subsection{Stress resultants and weak form of balance eqations}

From the molecule-to-molecule interaction potential $\Phi(x)=\Phi(\norm{x_{21}})$ the molecule-to-molecule interaction forces on molecule $1$ and molecule $2$ are defined as
\begin{align}
  \vf_{m1} = -  \pdiff{\Phi}{{\vx_{21}}} \quad \text{and} \quad \vf_{m2} = \pdiff{\Phi}{{\vx_{21}}}=-\vf_{m1}.
\end{align}
In a similar fashion, the differential force resultants between two differential volume elements $dV_1=dA_1ds_1$ and $dV_2=dA_2 ds_2$ at positions $\vx_1$ and $\vx_2$ within the cross-sections $A_1$ and $A_2$ is given by
\begin{subequations}
\begin{align}
  \dd \vf_{dV_1} &= - \rho_1(\mb{\xi}_1) \rho_2(\mb{\xi}_2) \pdiff{\Phi}{{\vx_{21}}} \dd V_2 \dd V_1\\
  \dd \vf_{dV_2} &= \rho_1(\mb{\xi}_1) \rho_2(\mb{\xi}_2) \pdiff{\Phi}{{\vx_{21}}} \dd V_2 \dd V_1 = -\dd \vf_{dV_1}
\end{align}
\end{subequations}
Integration over the two cross-sections yields the resulting differential force resultant between to cross-section slices (areas $A_1, A_2$, thicknesses $ds_1,ds_2$) with volumes $V_1=A_1ds_1$ and $V_2=A_2ds_2$:
\begin{subequations}
\label{f}
\begin{align}
  \dd \vf_{V_1} &= \int_{A_1} \int_{A_2} \dd \vf_{dV_1} =\underbrace{- \int_{A_1} \int_{A_2} \rho_1(\mb{\xi}_1) \rho_2(\mb{\xi}_2) \pdiff{\Phi}{{\vx_{21}}} \dd A_2 \dd A_1}_{:=\vf_1} ds_2 ds_1\\
  \dd \vf_{V_2} &= \int_{A_1} \int_{A_2} \dd \vf_{dV_2} =\underbrace{\int_{A_1} \int_{A_2} \rho_1(\mb{\xi}_1) \rho_2(\mb{\xi}_2) \pdiff{\Phi}{{\vx_{21}}} \dd A_2 \dd A_1}_{:=\vf_2} ds_2 ds_1 = -\dd \vf_{V_1}
\end{align}
\end{subequations}
Here, we have defined the \textit{section-section} interaction forces $\vf_1$ and $\vf_2$ of dimension "force per length square". Based on the differential forces $\dd \vf_{dV_1}$ and $\dd \vf_{dV_2}$, we can also define the differential moments with respect to the centroids $\vr_1$ and $\vr_2$ of two interaction cross-section slices with volumes $V_1$ and $V_2$:
\begin{subequations}
\label{m}
\begin{align}
  \dd \vm_{V_1|\vr_1} &= \int_{A_1} \int_{A_2}  \mb{\xi}_1 \times \dd \vf_{dV_1} = \underbrace{- \int_{A_1} \int_{A_2} \rho_1(\mb{\xi}_1) \rho_2(\mb{\xi}_2) \,\mb{\xi}_1 \times \pdiff{\Phi}{{\vx_{21}}} \dd A_2 \dd A_1}_{:=\vm_1} ds_2 ds_1\\
  \dd \vm_{V_2|\vr_2} &= \int_{A_1} \int_{A_2}  \mb{\xi}_2 \times \dd \vf_{dV_2} = \underbrace{\int_{A_1} \int_{A_2} \rho_1(\mb{\xi}_1) \rho_2(\mb{\xi}_2) \, \mb{\xi}_2 \times \pdiff{\Phi}{{\vx_{21}}} \dd A_2 \dd A_1}_{:=\vm_2} ds_2 ds_1
\end{align}
\end{subequations}
Here, we have defined the \textit{section-section} interaction moments $\vm_1$ and $\vm_2$ of dimension "moment per length square". With these definitions, the accumulated mechanical interaction between two cross-section slices with volumes $V_1=A_1 ds_1$ and $V_2=A_2 ds_2$ is given by the differential forces $\vf_1 ds_1 ds_2$ and $\vf_2 ds_1 ds_2$ acting at the cross-section centroids $\vr_1$ and $\vr_2$, respectively, together with the differential moments $\vm_1 ds_1 ds_2 $ and $\vm_2 ds_1 ds_2$. In a next step, the virtual work of the interaction forces $\dd \vf_{dV_1}$ acting on point $\vx_{1}$ within cross-section $1$ and $\dd \vf_{dV_2}$ acting on point $\vx_{2}$ within cross-section $2$ shall be formulated:
\begin{align}
\label{delta_W1}
  \delta W_{int} &= \int_{l_1} \int_{l_2} \int_{A_1} \int_{A_2} \left( \delta \vx_{1}^T \dd \vf_{dV_1} + \delta \vx_{2}^T \dd \vf_{dV_2} \right).
\end{align}
Employing~\eqref{simoreissner_x0} and~\eqref{largerotations_deltalambdaspatial}, the variations $\delta \vx_{1}$ and $\delta \vx_{2}$ of the cross-section positions $\vx_{1}$ and $\vx_{2}$ result in
\begin{align}
\label{delta_x12}
  \delta \vx_{i} &= \delta \vr_i +  \left(\xi_i^2  \delta \vtheta_i \times \vg_i^2 +\xi_i^3  \delta \vtheta_i \times \vg_i^3 \right)
  =\delta \vr_i +  \delta \vtheta_i \times \underbrace{\left(\xi_i^2   \vg_i^2 +\xi_i^3   \vg_i^3 \right)}_{\mb{\xi}_i}
  =\delta \vr_i - \vS(\mb{\xi}_i)\delta \vtheta_i   \quad \text{for} \quad i=1,2,
\end{align}
where $\delta \vr_1$ and $\delta \vr_2$ are the variations of the two cross-section centroid positions and $\delta \vtheta_1$ and $\delta \vtheta_2$ are the spin vectors associated with the cross-section triads $1$ and $2$, respectively. Inserting~\eqref{delta_x12} in~\eqref{delta_W1} yields:
\begin{align}
\begin{split}
\label{delta_W2}
\delta W_{int} &=\int_{l_1} \int_{l_2}  \delta \vr_1^T  \int_{A_1} \int_{A_2}  \dd \vf_{dV_1}
						     + \int_{l_1} \int_{l_2} \delta \vr_2^T \int_{A_1} \int_{A_2}   \dd \vf_{dV_2}\\
						     &+ \int_{l_1} \int_{l_2} \delta \vtheta_1^T \int_{A_1} \int_{A_2} \underbrace{\vS(\mb{\xi}_1) \dd \vf_{dV_1}}_{\mb{\xi}_1 \times \dd \vf_{dV_1}}
						     + \int_{l_1} \int_{l_2} \delta \vtheta_2^T \int_{A_1} \int_{A_2}  \underbrace{\vS(\mb{\xi}_2) \dd \vf_{dV_2}}_{\mb{\xi}_2 \times \dd \vf_{dV_2}}\\
   &= \int_{l_1} \int_{l_2} \left( \delta \vr_1^\text{T} \vf_1 + \delta \vr_2^\text{T} \vf_2 + \delta \vtheta_1^\text{T} \vm_1 + \delta \vtheta_2^\text{T} \vm_2 \right) ds_1 ds_2,
\end{split}
\end{align}
where the definitions of $\vf_1, \vf_2$ and $\vm_1,\vm_2$ according to~\eqref{f} and ~\eqref{m} have been utilized. As expected, the total virtual work is given by the products of the section interaction forces $\vf_i$ and moments $\vm_i$ with the associated work-conjugated virtual displacements $\delta \vr_i$ and rotations $\delta \vtheta_i$ integrated over the lengths $l_1$ and $l_2$ of both beams. In preparation to Section~\ref{sec:variational}, the virtual work shall be expressed as pure function of the stress resultants $\vf_2$ and $\vm_2$ acting on beam $2$. Inserting the equilibrium of forces and the equilibrium of moments with respect to the cross-section centroid position $\vr_1$ of beam $1$, i.e.
\begin{subequations}
\label{equilibrium}
\begin{align}
\vf_1&=-\vf_2, \label{equilibrium_1}\\
 \vm_1&=-\vm_2-(\vr_2-\vr_1) \times \vf_2 = -\vm_2-\vS(\vr_2-\vr_1) \vf_2, \label{equilibrium_2}
\end{align}
\end{subequations}
the virtual work expression~\eqref{delta_W2} can be reformulated as follows:
\begin{align}
\begin{split}
\label{delta_W3}
\delta W_{int} = \int_{l_1} \int_{l_2} \Big[  \underbrace{\left(\delta \vr_2^\text{T} - \delta \vr_1^\text{T} \!-\!  \delta \vtheta_1^T \! \vS[\left( \vr_2 - \vr_1 \right)] \right)}_{\delta_o \vr_{{2}{1}}^T} \vf_2 + \underbrace{\left( \delta \vtheta_2^\text{T} - \delta \vtheta_1^\text{T} \right)}_{\delta_o \vtheta_{{2}{1}}^T} \vm_2 \Big] ds_1 ds_2.
\end{split}
\end{align}
By defining material stress-resultants in analogy to the material deformation measures according to
\begin{subequations}
\label{material_stress_resultants}
\begin{align}
\vF_i&=\vLambda_1^T \vf_i \quad \text{for} \quad i=1,2 \label{material_stress_resultants_1},\\
\vM_i&=\vLambda_1^T \vm_i \quad \text{for} \quad i=1,2\label{material_stress_resultants_2},
\end{align}
\end{subequations}
also the material representation of the virtual work expression can formally be derived from~\eqref{delta_W3}:
\begin{align}
\begin{split}
\label{delta_W4}
\delta W_{int} = \int_{l_1} \int_{l_2} \Big[  \underbrace{\left(\delta \vr_2^\text{T} - \delta \vr_1^\text{T} \!-\!  \delta \vtheta_1^T \! \vS[\left( \vr_2 - \vr_1 \right)] \right) \vLambda_1}_{\delta \vR_{\mathbbm{2}\mathbbm{1}}^T} \vF_2 + \underbrace{\left( \delta \vtheta_2^\text{T} - \delta \vtheta_1^\text{T} \right) \vLambda_1}_{ \delta \vtheta_{\mathbbm{2}\mathbbm{1}}^T} \vM_2 \Big] ds_1 ds_2.
\end{split}
\end{align}
The prefactors of the spatial and material interaction forces in~\eqref{delta_W3} and~\eqref{delta_W4} can be identified as the objective (additive) variations of the translational deformation measures $\vr_{21}$ and $\vR_{\mathbbm{2}\mathbbm{1}}$ (see~\eqref{delta_o} and~\eqref{delta_material}). In the same fashion, the prefactors of the spatial and material interaction moments in~\eqref{delta_W3} and~\eqref{delta_W4} can be identified as the objective \textit{multiplicative} variations of the rotational deformation measures $\vpsi_{21}$ and $\vPsi_{\mathbbm{2}\mathbbm{1}}$. Thus, these generalized deformation measures indeed represent the kinematic quantities that are work-conjugated to the interaction forces and moments. In the next section, the spatial and material form of the virtual work will be derived on basis of a variational principle employing an interaction potential either as function of the spatial or of the material deformation measures, i.e., $\tilde{\pi}(\vr_{21},\vpsi_{21})$ or $\tilde{\pi}(\vR_{\mathbbm{2}\mathbbm{1}},\vPsi_{\mathbbm{2}\mathbbm{1}})$, as derived in Section~\eqref{sec:kinematics}.


\begin{center}
\begin{minipage}{0.9\textwidth}
{
\noindent \textbf{Remark:} Since the molecule-to-molecule interaction potential $\Phi(x)$ is only a function of the norm of $\vx_{21}$ or $\vX_{21}=\vLambda_1^T \vx_{21}$ with $x=\norm{\vx_{21}}=\norm{\vX_{21}}$, the result~\eqref{delta_W2} can equivalently be derived via variation of the total interaction potential\eqref{potential_1} employing 
$\delta \Phi(x)=(\partial \Phi / \partial x) \delta x = (\partial \Phi / \partial \vx_{21}) \delta \vx_{21} = (\partial \Phi / \partial \vX_{21}) \delta \vX_{21}$, i.e., no objective variation $\delta_o \vx_{21}$ has to be defined for the spatial object $\vx_{21}$ (the contribution from the variation of the moving base vectors cancels out because the potential is only a function of the norm of $\vx_{21}$). It will become clear in the next section that the situation is crucially different for the deformation measures $\vr_{21}$ and $\vpsi_{21}$. There, the objective variations~\eqref{delta_o} will be required in order to formulate an equivalent representation of the principle of virtual work~\eqref{delta_W2} that is based on a variational principle and a section-section interaction potential $\tilde{\pi}(\vr_{21},\vpsi_{21})$ as function of the spatial deformation measures $\vr_{21}$ and $\vpsi_{21}$.
}
\end{minipage}
\end{center}

\subsection{Variational problem statement based on a section-section interaction potential}
\label{sec:variational}

Let us consider the representation of the section-section interaction potential ${\bar{\pi}}(\vR_{\mathbbm{2}\mathbbm{1}},\vPsi_{\mathbbm{2}\mathbbm{1}})$ as function of the material deformation measures $\vR_{\mathbbm{2}\mathbbm{1}}$ and $\vPsi_{\mathbbm{2}\mathbbm{1}}$ according to~\eqref{barbar_pi}. Based on~\eqref{potential_1}, the variation 
of the total interaction potential can be written as:
\begin{align}
\label{delta_Pi_material}
\delta \Pi = \int_{l_1} \int_{l_2} \left[ \frac{\partial \bar{\pi}}{\partial \vR_{\mathbbm{2}\mathbbm{1}}} \delta \vR_{\mathbbm{2}\mathbbm{1}} +\frac{\partial \bar{\pi}}{\partial \vPsi_{\mathbbm{2}\mathbbm{1}}} \delta \vPsi_{\mathbbm{2}\mathbbm{1}} \right] \dd s_2 \dd s_1.
\end{align}
Transposing the inner products in~\eqref{delta_Pi_material} and inserting the variations $\delta \vR_{\mathbbm{2}\mathbbm{1}}$ and $\delta \vPsi_{\mathbbm{2}\mathbbm{1}}$ from~\eqref{delta_material} yields:
\begin{align}
\begin{split}
\label{delta_Pi_material_2}
\!\!\!\!\!\delta \Pi \!&=\!\!\! \int_{l_1} \! \int_{l_2} \! \left[ \left( \delta \vr_2^T \!-\! \delta \vr_1^T \!-\!  \delta \vtheta_1^T \! \vS[\left( \vr_2 - \vr_1 \right)] \right) \vLambda_1 \! \left(\frac{\partial \bar{\pi}}{\partial \vR_{\mathbbm{2}\mathbbm{1}}}\right)^{\!\!T} \!\!
+\left( \delta \vtheta_2^T \!-\! \delta \vtheta_1^T \right) \! \vLambda_1 \vT^T\!(\vPsi_{\mathbbm{2}\mathbbm{1}}) \!\! \left(\frac{\partial \bar{\pi}}{\partial \vPsi_{\mathbbm{2}\mathbbm{1}}}\right)^{\!\!T} \, \right] \dd s_2 \dd s_1 \!\!\!\!\! \\
								&=\!\!\! \int_{l_1} \! \int_{l_2} \! \Bigg[ \underbrace{\left( \delta \vr_2^T \!-\! \delta \vr_1^T \!-\!  \delta \vtheta_1^T \! \vS[\left( \vr_2 - \vr_1 \right)] \right) \vLambda_1}_{\delta \vR_{\mathbbm{2}\mathbbm{1}}^T} \! \left(\frac{\partial \bar{\pi}}{\partial \vR_{\mathbbm{2}\mathbbm{1}}}\right)^{\!\!T} \!\!
+\underbrace{\left( \delta \vtheta_2^T \!-\! \delta \vtheta_1^T \right) \! \vLambda_1}_{\delta \vtheta_{\mathbbm{2} \mathbbm{1}}^T} \left(\frac{\partial_m \bar{\pi}}{\partial_m \vtheta_{\mathbbm{2} \mathbbm{1}}}\right)^{\!\!T} \, \Bigg] \dd s_2 \dd s_1,\!\!\!\!\!
\end{split}
\end{align}
where from the first to the second line the definition of the multiplicative derivative according to~\eqref{multiplicative_derivative} has been identified. By considering the spatial representation of the section-section interaction potential ${\tilde{\pi}}(\vr_{21},\vpsi_{21})$ and employing the objective variations of the spatial deformation measures $\vr_{21}$ and $\vpsi_{21}$ according to~\eqref{delta_o}, the variation of the total interaction potential~\eqref{potential_1} can alternatively be written as 
\begin{align}
\begin{split}
\label{delta_Pi_spatial}
\!\!\!\!\!\delta \Pi \!&=\!\!\!\int_{l_1} \int_{l_2} \left[ \frac{\partial \tilde{\pi}}{\partial \vr_{21}} \delta_o \vr_{21} +\frac{\partial \tilde{\pi}}{\partial \vpsi_{21}} \delta_o \vpsi_{21} \right] \dd s_2 \dd s_1 \\ 
       &=\!\!\! \int_{l_1} \! \int_{l_2} \! \left[ \left( \delta \vr_2^T \!-\! \delta \vr_1^T \!-\!  \delta \vtheta_1^T \! \vS[\left( \vr_2 - \vr_1 \right)] \right) \! \left(\frac{\partial \tilde{\pi}}{\partial \vr_{21}}\right)^{\!\!T} \!\!
+\left( \delta \vtheta_2^T \!-\! \delta \vtheta_1^T \right) \vT^T\!(\vpsi_{21}) \!\! \left(\frac{\partial \tilde{\pi}}{\partial \vpsi_{21}}\right)^{\!\!T} \, \right] \dd s_2 \dd s_1 \!\!\!\!\! \\
								&=\!\!\! \int_{l_1} \! \int_{l_2} \! \Bigg[ \underbrace{\left( \delta \vr_2^T \!-\! \delta \vr_1^T \!-\!  \delta \vtheta_1^T \! \vS[\left( \vr_2 - \vr_1 \right)] \right)}_{\delta_o \vr_{21}^T} \! \left(\frac{\partial \tilde{\pi}}{\partial \vr_{21}}\right)^{\!\!T} \!\!
+\underbrace{\left( \delta \vtheta_2^T \!-\! \delta \vtheta_1^T \right)}_{\delta_o \vtheta_{21}^T} \! \left(\frac{\partial_m \tilde{\pi}}{\partial_m \vtheta_{21}}\right)^{\!\!T} \, \Bigg] \dd s_2 \dd s_1,\!\!\!\!\!
\end{split}
\end{align}
By requiring $\delta W_{int} = \delta \Pi$ for the spatial and material virtual work expressions~\eqref{delta_W3} and~\eqref{delta_W4} and the spatial and material variations~\eqref{delta_Pi_spatial} and~\eqref{delta_Pi_material_2}, the spatial and material stress-resultants can be identified as:
\begin{subequations}
\label{stress_resultants_via_potential}
\begin{align}
\vf_2&=\left(\frac{\partial \tilde{\pi}}{\partial \vr_{21}}\right)^{\!\!T}, \quad \quad \vm_2=\left(\frac{\partial_m \tilde{\pi}}{\partial_m \vtheta_{21}}\right)^{\!\!T}=\vT^T\!(\vpsi_{21})\left(\frac{\partial \tilde{\pi}}{\partial \vpsi_{21}}\right)^{\!\!T}, \label{stress_resultants_via_potential_1}\\
\vF_2&=\left(\frac{\partial \bar{\pi}}{\partial \vR_{\mathbbm{2}\mathbbm{1}}}\right)^{\!\!T}, \quad \quad \vM_2=\left(\frac{\partial_m \bar{\pi}}{\partial_m \vtheta_{\mathbbm{2}\mathbbm{1}}}\right)^{\!\!T}=\vT^T\!(\vPsi_{\mathbbm{2}\mathbbm{1}})\left(\frac{\partial \tilde{\pi}}{\partial \vPsi_{\mathbbm{2}\mathbbm{1}}}\right)^{\!\!T}. \label{stress_resultants_via_potential_2}
\end{align}
\end{subequations}
Thus, the important result of this section is that the interaction between beams due to molecular interaction potentials and the resulting contributions to (the weak or strong form of) the mechanical balance equations can be formulated on the basis of a variational principle considering section-section interaction potentials ${\tilde{\pi}}(\vr_{21},\vpsi_{21})$ or ${\bar{\pi}}(\vR_{\mathbbm{2}\mathbbm{1}},\vPsi_{\mathbbm{2}\mathbbm{1}})$ that are a pure function of a minimal set of spatial or material kinematic quantities, denoted as generalized deformation measures. Importantly, while the interaction forces are defined as \textit{additive derivative} (i.e., partial derivative) of the section-section interaction potential with respect to the translational deformation measure, the interaction moments are defined as \textit{multiplicative derivative} of the section-section interaction potential with respect to the rotational deformation measure. 

For given molecule-to-molecule interaction potentials $\Phi(x)$, the most obvious way of deriving these section-section interaction potentials is to perform the two cross-section integrals in~\eqref{potential_1} analytically, by setting either $x=||\vx_{21}||$ or $x=||\vX_{21}||$. Often, however, an analytical solution of these integrals (or at least of a reasonable approximation) is not possible in closed form. Alternatively, section-section interaction potentials as functions of the deformation measures $(\vr_{21},\vpsi_{21})$ or $(\vR_{\mathbbm{2}\mathbbm{1}},\vPsi_{\mathbbm{2}\mathbbm{1}})$ might be found via curve fitting based on values ${\tilde{\pi}}^n(\vr_{21}^n,\vpsi_{21}^n)$ or ${\bar{\pi}}^n(\vR_{\mathbbm{2}\mathbbm{1}}^n,\vPsi_{\mathbbm{2}\mathbbm{1}}^n)$ at discrete sampling points found experimentally or via numerical integration of the cross-section integrals.

\begin{center}
\begin{minipage}{0.9\textwidth}
{
\noindent \textbf{Remark:} Throughout this work, the cross-section triad $\vLambda_1$ has been considered as reference triad and pull-forward operator. Equivalently, material deformation measures could have been defined using $\vLambda_2$ as push-forward operator. Of course, the final virtual work expressions, i.e., the resulting weak form of the balance equations would be equivalent, with the main difference that in this case the stress-resultants on beam $1$, i.e., $\vf_1,\vm_1,\vF_1,\vM_1$ would naturally occur in the variational problem statement according to~\eqref{stress_resultants_via_potential}.
}
\end{minipage}
\end{center}

\subsection{Objectivity}
\label{sec:objectivity}

To verify objectivity, a rigid body rotation with rotation tensor $\vR^*$ is superimposed according to
\begin{align}
\label{objectivty_!}
 \vr_{1}^* = \vR^* \vr_{1}, \quad \vr_{2}^* = \vR^* \vr_{2},\quad \vLambda_1^*= \vR^*\vLambda_1 \quad \vLambda_2^*= \vR^*\vLambda_2,
\end{align}
where the superscript $(.)^*$ represents the rotated configuration. Based on equations~\eqref{d_and_psi},~\eqref{D_and_Psi} and~\eqref{A_Lambda1}, it can be verified 
that the spatial and material deformation measures of the rotated configuration read
\begin{align}
\label{objectivty_!}
 \vr_{21}^* = \vR^* \vr_{21}, \quad \vpsi_{21}^* = \vR^* \vpsi_{21},\quad \vR_{\mathbbm{2}\mathbbm{1}}^*= \vR_{\mathbbm{2}\mathbbm{1}} \quad \vPsi_{\mathbbm{2}\mathbbm{1}}^*= \vPsi_{\mathbbm{2}\mathbbm{1}},
\end{align}
which already proofs objectivity of the spatial and material deformation measures. As a simple counter-example, it can easily be verified that an alternative spatial rotational deformation measure formulated as the difference of the two triad orientation angles according to
\begin{align}
\label{objectivty_!}
\tilde{\vpsi}_{21}:=\vpsi_{2}-\vpsi_{1}
\end{align}
would not fulfill spatial objectivity in general since $\tilde{\vpsi}_{21}^*=\vpsi_{2}^*-\vpsi_{1}^*= \text{rv}(\vR^*\vLambda_2)-\text{rv}(\vR^*\vLambda_1) \neq \vR^* \tilde{\vpsi}_{21}$, which becomes obvious by the following reformulation $\vR^* \tilde{\vpsi}_{21}= \vR^*\vpsi_{2} - \vR^*\vpsi_{1} = \text{rv}(\vR^*\vLambda_2 \vR^{* T} )-\text{rv}(\vR^*\vLambda_1\vR^{* T})$.

\section{Special SSIP laws with high practical relevance}
\label{sec:specialSSIP}

In this section, four special SSIP laws with high practical relevance will be presented, namely SSIPs with distinct reference configuration, hyperelastic stored-energy functions in the framework of the geometrically exact beam theory, quadratic forms as simplest case of SSIPs, and SSIPs as penalty or Lagrange multiplier potentials to enforce general translational and rotational constraints.

\subsection{SSIPs with distinct reference configuration}
\label{sec:distinct}

If there is a distinct reference configuration $\vR_{\mathbbm{2}\mathbbm{1}}^0,\vPsi_{\mathbbm{2}\mathbbm{1}}^0$, at which the section-section interaction potential takes on a prescribed value, e.g., ${\bar{\pi}}(\vR_{\mathbbm{2}\mathbbm{1}}^0,\vPsi_{\mathbbm{2}\mathbbm{1}}^0)=0$, the original deformation measures can be replaced by $\hat{\vR}_{\mathbbm{2}\mathbbm{1}}=\vR_{\mathbbm{2}\mathbbm{1}}-\vR_{\mathbbm{2}\mathbbm{1}}^0$ and $\hat{\vPsi}_{\mathbbm{2}\mathbbm{1}}=\vPsi_{\mathbbm{2}\mathbbm{1}}-\vPsi_{\mathbbm{2}\mathbbm{1}}^0$  in order to end up with a simpler expression for the section-section interaction potential. In a similar fashion, also the spatial deformation measures can be replaced by $\hat{\vr}_{{2}{1}}=\vLambda_1 \hat{\vR}_{\mathbbm{2}\mathbbm{1}}=\vr_{{2}{1}}-\vLambda_1\vR_{\mathbbm{2}\mathbbm{1}}^0$ and $\hat{\vpsi}_{\mathbbm{2}\mathbbm{1}}=\vLambda_1 \hat{\vPsi}_{\mathbbm{2}\mathbbm{1}}=\vpsi_{{2}{1}}-\vLambda_1\vPsi_{\mathbbm{2}\mathbbm{1}}^0$. Of course, the additional constants do not change the variational calculus and the structure of the resulting equations as presented above, i.e., $\delta \hat{\vR}_{\mathbbm{2}\mathbbm{1}} =  \delta \vR_{\mathbbm{2}\mathbbm{1}}$ and $\delta \hat{\vPsi}_{\mathbbm{2}\mathbbm{1}} = \delta \vPsi_{\mathbbm{2}\mathbbm{1}}$ as well as $\delta_o \hat{\vr}_{{2}{1}} =  \delta_o \vr_{{2}{1}}$ and $\delta_o \hat{\vpsi}_{{2}{1}} = \delta_o \vpsi_{{2}{1}}$. In the following sections, this more general representation of SSIP laws will be considered.

\begin{center}
\begin{minipage}{0.9\textwidth}
{
\noindent \textbf{Remark:} Using~\eqref{obj_var_general}, the objective variations of $\hat{\vr}_{{2}{1}}$ and $\hat{\vpsi}_{{2}{1}}$ as stated above can be verified:
\begin{align}
\begin{split}
\label{verification_variation_hat}
 \delta_o \hat{\vr}_{{2}{1}}&=\delta \hat{\vr}_{{2}{1}} - \vS(\delta \vtheta_1) \hat{\vr}_{{2}{1}}
 =\delta (\vr_{{2}{1}}-\vLambda_1\vR_{\mathbbm{2}\mathbbm{1}}^0) - \vS(\delta \vtheta_1) (\vr_{{2}{1}}-\vLambda_1\vR_{\mathbbm{2}\mathbbm{1}}^0)  \\
 &= \delta_o \vr_{{2}{1}} - \vS(\delta \vtheta_1) \vLambda_1\vR_{\mathbbm{2}\mathbbm{1}}^0 + \vS(\delta \vtheta_1) \vLambda_1\vR_{\mathbbm{2}\mathbbm{1}}^0
 = \delta_o \vr_{{2}{1}} \checkmark \\
 \delta_o \hat{\vpsi}_{{2}{1}}&=\delta \hat{\vpsi}_{{2}{1}} - \vS(\delta \vtheta_1) \hat{\vpsi}_{{2}{1}}
 =\delta (\vpsi_{{2}{1}}-\vLambda_1\vPsi_{\mathbbm{2}\mathbbm{1}}^0) - \vS(\delta \vtheta_1) (\vpsi_{{2}{1}}-\vLambda_1\vPsi_{\mathbbm{2}\mathbbm{1}}^0)  \\
 &= \delta_o \vpsi_{{2}{1}} - \vS(\delta \vtheta_1) \vLambda_1\vPsi_{\mathbbm{2}\mathbbm{1}}^0 + \vS(\delta \vtheta_1) \vLambda_1\vPsi_{\mathbbm{2}\mathbbm{1}}^0 = \delta_o \vpsi_{{2}{1}} \checkmark
 \end{split}
\end{align}
}
\end{minipage}
\end{center}

\subsection{Hyperelastic stored-energy function as asymptotic limiting case of SSIPs}

In the special case that the considered SSIP represents interactions between pairs of molecules within \textit{one} beam, i.e., internal elastic forces, the interacting cross-sections $A_1$ and $A_2$ as well as the two outer integration loops along $l_1$ and $l_2$ in~\eqref{potential_1} refer to one and the same beam. Thus, a given cross-section of a beam interacts with all other cross-sections within this beam. However, since these interactions are typically very short-ranged, the boundaries of the integration over the coordinate ${s}_2$ can be adapted to $s_1- \Delta$ and $s_1+ \Delta$. This leads to the following expression for the total interaction potential:
\begin{align}
\label{SSIP_SimoReissner_1}
  \Pi = \int \limits_{0}^{l_1} \int \limits_{s_1- \Delta}^{s_1+\Delta} \bar{\pi}(\underbrace{\vR_{\mathbbm{2}\mathbbm{1}}-\vR_{\mathbbm{2}\mathbbm{1}}^0}_{\hat{\vR}_{\mathbbm{2}\mathbbm{1}}},\underbrace{\vPsi_{\mathbbm{2}\mathbbm{1}}-\vPsi_{\mathbbm{2}\mathbbm{1}}^0}_{\hat{\vPsi}_{\mathbbm{2}\mathbbm{1}}}) \dd {s}_2 \dd s_1.
\end{align}
Here, the reference state $\vR_{\mathbbm{2}\mathbbm{1}}^0,\vPsi_{\mathbbm{2}\mathbbm{1}}^0$ is considered as stress-free, i.e., $\bar{\pi}=0$ for $\vR_{\mathbbm{2}\mathbbm{1}}=\vR_{\mathbbm{2}\mathbbm{1}}^0$ and $\vPsi_{\mathbbm{2}\mathbbm{1}}=\vPsi_{\mathbbm{2}\mathbbm{1}}^0$ (see Section~\ref{sec:distinct} above). Due to the short-range nature of interactions between the cross-sections at ${s}_1$ and $s_2={s}_1+\tilde{s}_2$ ($|\tilde{s}_2| \ll l_2$), the kinematics of the second cross-section are approximated based on a first-order Taylor series expansion at ${s}_1$. Thus, the translational deformation measure can be reformulated as follows:
\begin{align}
\begin{split}
\label{SSIP_SimoReissner_2}
  \vR_{\mathbbm{2}\mathbbm{1}} - \vR_{\mathbbm{2}\mathbbm{1}}^0 = & \, \vLambda_1^T(s_1)  [\vr_1(s_1+\tilde{s}_2)-\vr_1(s_1)] - \vLambda_1^{0T}(s_1)  [\vr_1^0(s_1+\tilde{s}_2)-\vr_1^0(s_1)] \\
 = &
\,  \,\,\,\,\,\,\, \vLambda_1^T(s_1)  [\vr_1(s_1) + \vr_1^{\prime}(s_1) \tilde{s}_2 + \mathcal O(\tilde{s}_2^2) -\vr_1(s_1)] \\
 &\, - \vLambda_1^{0T}(s_1)  [\vr_1^0(s_1) + \vr_1^{0 \prime}(s_1) \tilde{s}_2 + \mathcal O(\tilde{s}_2^2) -\vr_1^0(s_1)] \\
 \approx \, & [\vLambda_1^T(s_1)  \vr_1^{\prime}(s_1) - \vLambda_1^{0T}(s_1)  \vr_1^{0 \prime}(s_1)]\tilde{s}_2  \\
 = & \, 
\underbrace{ [\vLambda_1^T(s_1)  \vr_1^{\prime}(s_1) - \ve^1]}_{=\boldsymbol{\Gamma}(s_1)}\tilde{s}_2.
\end{split}
\end{align}
Here, the relations $\vr_1^{0 \prime}=\vg_1^{10}$ and $\vLambda_1^{0}=\vg_1^{i0} \otimes \ve^i$ have been used in the last reformulation step. In a similar fashion, also the rotational deformation measures are reformulated:
\begin{align}
\begin{split}
\label{SSIP_SimoReissner_3}
 \vPsi_{\mathbbm{2}\mathbbm{1}}-\vPsi_{\mathbbm{2}\mathbbm{1}}^0&= \text{rv}\{\vLambda_1^T(s_1)  \vLambda_1(s_1+\tilde{s}_2)\} - \text{rv}\{\vLambda_1^{0T}(s_1)  \vLambda_1^0(s_1+\tilde{s}_2)\} \\
 & =
 \text{rv}\{\vLambda_1^T(s_1)  [\vLambda_1(s_1) + \vLambda_1^{\prime}(s_1) \tilde{s}_2 + \mathcal O(\tilde{s}_2^2)]\} -\text{rv}\{\vLambda_1^{0T}(s_1)  [\vLambda_1^{0}(s_1) + \vLambda_1^{0 \prime}(s_1) \tilde{s}_2 + \mathcal O(\tilde{s}_2^2)]\} \\
 & = 
\text{rv}\{ [\vI + \vLambda_1^T(s_1) \vLambda_1^{\prime}(s_1) \tilde{s}_2 + \mathcal O(\tilde{s}_2^2) ]\} - \text{rv}\{[\vI + \vLambda_1^{0T}(s_1) \vLambda_1^{0 \prime}(s_1) \tilde{s}_2 + \mathcal O(\tilde{s}_2^2)]\} \\
  & = 
 \text{rv}\{[\vI + \vS(\tilde{s}_2 \vK(s_1)) + \mathcal O(\tilde{s}_2^2) ]\} - \text{rv}\{[\vI + \vS(\tilde{s}_2 \vK^0(s_1)) + \mathcal O(\tilde{s}_2^2)]\} \\
  & = 
 \text{rv}\{[\exp\{ \vS(\tilde{s}_2 \vK(s_1))\}  + \mathcal O(\tilde{s}_2^2) ]\} - \text{rv}\{[\exp\{ \vS(\tilde{s}_2 \vK^0(s_1))\}  + \mathcal O(\tilde{s}_2^2)]\}\\
  & \approx 
 \text{rv}\{\exp\{ \vS(\tilde{s}_2 \vK(s_1))\}\} - \text{rv}\{\exp\{ \vS(\tilde{s}_2 \vK^0(s_1))\} \} \\
 & = 
\underbrace{ [\vK(s_1)  - \vK^0(s_1)] }_{=\boldsymbol{\Omega}(s_1)}\tilde{s}_2,
\end{split}
\end{align}
From the fourth to the fifth line, a first-order Taylor series expansion of the Rodrigues formula according to $\exp\{ \vS(\tilde{s}_2 \vK)\}= \vI + \vS(\tilde{s}_2 \vK) + \mathcal O(\tilde{s}_2^2)$ has been employed, which can be directly verified by expanding~\eqref{largerotations_rotrigues} with respect to $\tilde{s}_2$ using $\boldsymbol{\psi}=\tilde{s}_2 \vK$. Importantly, according to~\eqref{SSIP_SimoReissner_2}~\eqref{SSIP_SimoReissner_3} the proposed generalized SSIP deformation measures asymptotically converge towards the deformation measures $\boldsymbol{\Gamma}$ and $\boldsymbol{\Omega}$ of the geometrically exact beam theory in the limiting case of small relative positions and rotations between the interacting cross-sections. Inserting these relations into~\eqref{SSIP_SimoReissner_1} and integrating along $\tilde{s}_2$ yields for the total interaction potential
\begin{align}
\begin{split}
\label{SSIP_SimoReissner_4}
  \Pi & \approx \int \limits_{0}^{l_1} \int \limits_{- \Delta}^{\Delta} \bar{\pi}(\boldsymbol{\Gamma}(s_1)\tilde{s}_2,\boldsymbol{\Omega}(s_1)\tilde{s}_2) \dd \tilde{s}_2 \dd s_1 \\
  & = \int \limits_{0}^{l_1} \bar{\pi}_{\tilde{s}_2}(\boldsymbol{\Gamma}(s_1),\boldsymbol{\Omega}(s_1)) \dd s_1,
\end{split}
\end{align}
where $\bar{\pi}_{\tilde{s}_2}$ represents the antiderivative of $\bar{\pi}$ with respect to $\tilde{s}_2$ evaluated at the boundaries $\tilde{s}_2= \pm \Delta$. Comparison of~\eqref{SSIP_SimoReissner_4} with~\eqref{potential_energy} reveals, that the length-specific interaction potential $\bar{\pi}_{\tilde{s}_2}(\mb{\Gamma},\mb{\Omega})$, i.e., a section-section interaction potential $\bar{\pi}$ once integrated along the length $l_1$, can be identified as the hyperelastic stored-energy function $\bar{\pi}_{int}(\mb{\Gamma},\mb{\Omega})$ associated with the internal forces and moments of the geometrically exact beam theory. Based on the relations above, similar identities can be derived also for the stress resultants, the objective variations of the deformation measures and the weak form of the balance equations of both theories when considering the limiting case of small relative positions and rotations between the interacting cross-sections. In other words, the proposed theory of generalized section-to-section interaction potentials can be interpreted as a generalization of the geometrically exact Simo-Reissner beam theory in the sense that interactions between cross-sections with arbitrary relative positions and rotations are considered.

\begin{center}
\begin{minipage}{0.9\textwidth}
{
\noindent \textbf{Remark:} Based on the derivations above, the following alternative definitions of the spatial and material curvature vector as introduced in equation~\eqref{simoreissner_spatialcurvature} can be stated: 
\begin{align*}
\begin{split}
 \mb{k} &= \lim_{\Delta s \rightarrow 0} \frac{\vpsi_{{2}{1}}}{\Delta s}  \quad \text{with} \quad \vpsi_{{2}{1}}= \text{rv}\{  \vLambda(s+\Delta s) \vLambda^T(s) \} 
 \quad \text{or} \quad \vLambda(s+\Delta s)=\exp\{ \vS( \vpsi_{{2}{1}}) \} \vLambda(s) \\
 \mb{K} &=  \lim_{\Delta s \rightarrow 0} \frac{\vPsi_{\mathbbm{2}\mathbbm{1}}}{\Delta s}  \quad \text{with} \quad \vPsi_{\mathbbm{2}\mathbbm{1}}= \text{rv}\{  \vLambda^T(s) \vLambda(s+\Delta s)  \}  \quad \text{or} \quad \vLambda(s+\Delta s)= \vLambda(s) \exp\{ \vS( \vPsi_{\mathbbm{2}\mathbbm{1}}) \}
\end{split}
\end{align*}
}
\end{minipage}
\end{center}

\subsection{SSIP laws in quadratic form}

As an illustrative example, the simple case of a quadratic SSIP $\tilde{\pi}$ shall be considered according to
\begin{align}
\label{example_spatial_potential}
\!\!\! \tilde{\pi}(\hat{\vr}_{21},\hat{\vpsi}_{21})=\frac{1}{2}\hat{\vr}_{21}^T\vc_{\vr}\hat{\vr}_{21} + \frac{1}{2}\hat{\vpsi}_{21}^T\vc_{\vpsi}\hat{\vpsi}_{21} \,\, \rightarrow \,\,
\vf_2=\vc_{\vr}\hat{\vr}_{21}, \quad \vm_2=\vT^T\!(\vpsi_{21}) \vc_{\vpsi}\hat{\vpsi}_{21},\!\!\!
\end{align}
with the constant symmetric positive definite material tensors $c_{\vr}$ and $\vc_{\vpsi}$. Please note, that the transformation matrix $\vT^T\!(\vpsi_{21})$ in $\vm_2$ is a result of the objective variation $\delta_o \hat{\vpsi}_{{2}{1}} = \delta_o \vpsi_{{2}{1}} = \vT(\vpsi_{21}) \left( \delta \vtheta_2 - \delta \vtheta_1 \right)$ according to~\eqref{delta_o} and~\eqref{delta_Pi_spatial}. Thus, its argument is $\vpsi_{21}$ and \textit{not} $\hat{\vpsi}_{21}$. Replacing the spatial by the material potential according to $\bar{\pi}(\hat{\vR}_{\mathbbm{2}\mathbbm{1}},\hat{\vPsi}_{\mathbbm{2}\mathbbm{1}})=\tilde{\pi}(\vLambda_1 \hat{\vR}_{\mathbbm{2}\mathbbm{1}},\vLambda_1 \hat{\vPsi}_{\mathbbm{2}\mathbbm{1}})$ yields the material representation of the SSIP:
\begin{align}
\label{example_material_potential}
\!\!\! \bar{\pi}(\hat{\vR}_{\mathbbm{2}\mathbbm{1}},\hat{\vPsi}_{\mathbbm{2}\mathbbm{1}})=\frac{1}{2}\hat{\vR}_{\mathbbm{2}\mathbbm{1}}^T\vC_{\vR}v\hat{\vR}_{\mathbbm{2}\mathbbm{1}} + \frac{1}{2}\hat{\vPsi}_{\mathbbm{2}\mathbbm{1}}^T\vC_{\vPsi}\hat{\vPsi}_{\mathbbm{2}\mathbbm{1}} \,\, \rightarrow \,\, 
\vF_2=\vC_{\vR}\hat{\vR}_{\mathbbm{2}\mathbbm{1}}, \quad \vM_2=\vT^T\!({\vPsi}_{\mathbbm{2}\mathbbm{1}}) \vC_{\vPsi}\hat{\vPsi}_{\mathbbm{2}\mathbbm{1}}.\!\!\!
\end{align}
with $C_{\vR}=\vLambda_1^T c_{\vr} \vLambda_1$ and $C_{\vPsi}=\vLambda_1^T c_{\vpsi} \vLambda_1$. Exemplarily, a coordinate representation in the frame $\vg_1^i$ reads:
\begin{align}
\label{constants}
\vc_{\vr}=c_{\vr}^{ij} \vg_i \otimes \vg_j^T, \,\,\,  \vc_{\vpsi}=c_{\vpsi}^{ij} \vg_i \otimes \vg_j^T, \quad \quad
\vC_{\vR}=c_{\vr}^{ij} \ve_i \otimes \ve_j^T, \,\,\,  \vC_{\vPsi}=c_{\vpsi}^{ij} \ve_i \otimes \ve_j^T.
\end{align}
The interaction forces and moments in~\eqref{example_spatial_potential} and~\eqref{example_material_potential} have been calculated according to~\eqref{stress_resultants_via_potential}. Interestingly, the interaction moments resulting from a quadratic potential according to~\eqref{example_spatial_potential} and~\eqref{example_material_potential} are a nonlinear function of the rotational deformation measures $\hat{\vpsi}_{21}$ and $\hat{\vPsi}_{\mathbbm{2}\mathbbm{1}}$ in general, which is a direct consequence of the transformations $\vT\!(\vpsi_{21})$ and $\vT\!(\vPsi_{\mathbbm{2}\mathbbm{1}})$ between additive and multiplicative rotation increments. This is an important difference to the internal stress resultants~\eqref{storedenergyfunction_stressresultants} of the geometrically exact beam theory, where the relative rotations between the interacting cross-sections are infinitesimally small. In this limit of small relative rotations $\vPsi_{\mathbbm{2}\mathbbm{1}} \rightarrow$ \textbf{0}, the transformation matrices tend to identity, i.e., $\vT\!(\vpsi_{21})\rightarrow \vI_3, \, \vT\!(\vPsi_{\mathbbm{2}\mathbbm{1}})\rightarrow \vI_3$, and the relation between interaction moments and rotational deformation measures becomes linear.

\subsection{SSIPs for constraint enforcement}
In this section it will be demonstrated how SSIP laws can be formulated as penalty or Lagrange multiplier potentials to enforce positional and rotational constraints of the following from:
\begin{align}
\begin{split}
\label{spatial_penalty_potential0}
\vr_{21}&\dot{=}\vr_{21}^0 \quad  {and} \quad \vpsi_{21}\dot{=}\vpsi_{21}^0 \quad \quad \text{or} \quad \quad \hat{\vr}_{21}=\vr_{21}-\vr_{21}^0 \dot{=}0  \quad \text{and} \quad \hat{\vpsi}_{21}=\vpsi_{21}-\vpsi_{21}^0 \dot{=}0,\\
\vR_{\mathbbm{2}\mathbbm{1}}&\dot{=}\vR_{\mathbbm{2}\mathbbm{1}}^0 \quad {and} \quad
\vPsi_{\mathbbm{2}\mathbbm{1}}\dot{=}\vPsi_{\mathbbm{2}\mathbbm{1}}^0 
\quad \quad \text{or} \quad \quad \hat{\vR}_{\mathbbm{2}\mathbbm{1}}=\vR_{\mathbbm{2}\mathbbm{1}}-\vR_{\mathbbm{2}\mathbbm{1}}^0 \dot{=}0  \quad \text{and} \quad \hat{\vPsi}_{\mathbbm{2}\mathbbm{1}}=\vPsi_{\mathbbm{2}\mathbbm{1}}-\vPsi_{\mathbbm{2}\mathbbm{1}}^0 \dot{=}0.
\end{split}
\end{align}
It is emphasized that the presented procedure is not limited to the coupling of two beam cross-sections. Instead, it is valid for general mechanical problems involving the coupling of positions and rotations. Starting from~\eqref{example_spatial_potential} and~\eqref{example_material_potential}, a penalty potential is defined by simply choosing the material constants as $c_{\vr}^{ij}= \epsilon_{\vr} \delta^{ij}$ and $c_{\vpsi}^{ij}= \epsilon_{\vpsi} \delta^{ij}$ with the Kronecker delta $\delta^{ij}$ as well as the translational and rotational penalty parameters  $\epsilon_{\vr}$ and $\epsilon_{\vpsi}$. In this case, the material tensors simplify to $\vc_{\vr}=\vC_{\vR}=\epsilon_{\vr} \vI$ and $\vc_{\vpsi}=\vC_{\vPsi}=\epsilon_{\vpsi} \vI$:
\begin{align}
\begin{split}
\label{spatial_penalty_potential}
\tilde{\pi}_{\epsilon}(\hat{\vr}_{21},\hat{\vpsi}_{21})& =\frac{1}{2}\epsilon_{\vr} \hat{\vr}_{21}^T \hat{\vr}_{21} + \frac{1}{2}\epsilon_{\vpsi} \hat{\vpsi}_{21}^T \hat{\vpsi}_{21} \,\, \rightarrow \,\,
\vf_2= \epsilon_{\vr} \hat{\vr}_{21}, \quad \vm_2= \vT^T\!({\vpsi}_{{2}{1}}) \epsilon_{\vpsi} \hat{\vpsi}_{21}, \\
\bar{\pi}_{\epsilon}(\hat{\vR}_{\mathbbm{2}\mathbbm{1}},\hat{\vPsi}_{\mathbbm{2}\mathbbm{1}})&=\frac{1}{2}\epsilon_{\vr} \hat{\vR}_{\mathbbm{2}\mathbbm{1}}^T\hat{\vR}_{\mathbbm{2}\mathbbm{1}} + \frac{1}{2}\epsilon_{\vpsi} \hat{\vPsi}_{\mathbbm{2}\mathbbm{1}}^T\hat{\vPsi}_{\mathbbm{2}\mathbbm{1}} \,\, \rightarrow \,\, 
\vF_2=\epsilon_{\vr} \hat{\vR}_{\mathbbm{2}\mathbbm{1}}, \quad \vM_2= \vT^T\!({\vPsi}_{\mathbbm{2}\mathbbm{1}}) \epsilon_{\vpsi} \hat{\vPsi}_{\mathbbm{2}\mathbbm{1}}.
\end{split}
\end{align}
Eventually, inserting the penalty forces and moments into~\eqref{delta_Pi_material_2} and~\eqref{delta_Pi_spatial}, the virtual work contribution due to a penalty potential can be stated. Without loss of generality, the constraint at one local point is considered in the following, i.e., the integration in beam length direction is omitted to shorten notation.
\begin{align}
\begin{split}
\label{penalty_potential_variation}
\!\!\!\!\!\delta \pi_{\epsilon} 
\!&=   
\underbrace{( \delta \vr_2^T \!-\! \delta \vr_1^T \!-\!  \delta \vtheta_1^T \! \vS[\left( \vr_2 - \vr_1 )] \right) }_{\delta_o \vr_{{2}{1}}^T} \underbrace{\epsilon_{\vr} \hat{\vr}_{{2}{1}}}_{\vf_2}
+ \underbrace{( \delta \vtheta_2^T \!-\! \delta \vtheta_1^T ) }_{\delta_o \vtheta_{{2} {1}}^T} 
\underbrace{ \vT^T\!({\vpsi}_{{2}{1}}) \epsilon_{\vpsi} \hat{\vpsi}_{{2}{1}}}_{\vm_2} \\
&=   \underbrace{( \delta \vr_2^T \!-\! \delta \vr_1^T \!-\!  \delta \vtheta_1^T \! \vS[\left( \vr_2 - \vr_1 )] \right) \vLambda_1}_{\delta \vR_{\mathbbm{2}\mathbbm{1}}^T} \underbrace{\epsilon_{\vr} \hat{\vR}_{\mathbbm{2}\mathbbm{1}}}_{\vF_2}
+ \underbrace{( \delta \vtheta_2^T \!-\! \delta \vtheta_1^T ) \! \vLambda_1}_{\delta \vtheta_{\mathbbm{2} \mathbbm{1}}^T} 
\underbrace{\vT^T\!({\vPsi}_{\mathbbm{2}\mathbbm{1}}) \epsilon_{\vpsi}  \hat{\vPsi}_{\mathbbm{2}\mathbbm{1}}}_{\vM_2}
\end{split}
\end{align}
Since a rotation vector $\vpsi$ (and any vector parallel to $\vpsi$) is an eigenvector of the associated transformation matrix, i.e., $\vT\!(\vpsi) \vpsi = \vT^T\!(\vpsi) \vpsi = \vpsi$ (see~\eqref{A_T3}), the penalty moments simplify for the special case $\vPsi_{\mathbbm{2}\mathbbm{1}}^0=\mb{0}$, i.e., $\hat{\vPsi}_{\mathbbm{2}\mathbbm{1}}=\vPsi_{\mathbbm{2}\mathbbm{1}}$ and $\hat{\vpsi}_{{2}{1}}=\vpsi_{{2}{1}}$, to $\vm_2=\epsilon_{\vpsi} \hat{\vpsi}_{21}$ and $\vM_2=\epsilon_{\vpsi} \hat{\vPsi}_{\mathbbm{2}\mathbbm{1}}$ (see~\cite{steinbrecher2021}). As alternative to the penalty approach, the constraints~\eqref{spatial_penalty_potential0} may be enforced by a Lagrange multiplier potential according to:
\begin{align}
\begin{split}
\label{spatial_LM_potential}
\tilde{\pi}_{\vlambda}(\hat{\vr}_{21},\hat{\vpsi}_{21})& =\vlambda_{{\vr}}^T \hat{\vr}_{21} + \vlambda_{{\vpsi}}^T \hat{\vpsi}_{21}, \\
\bar{\pi}_{\vlambda}(\hat{\vR}_{\mathbbm{2}\mathbbm{1}},\hat{\vPsi}_{\mathbbm{2}\mathbbm{1}})&= \vlambda_{\vR}^T \hat{\vR}_{\mathbbm{2}\mathbbm{1}} + \vlambda_{{\vPsi}}^T \hat{\vPsi}_{\mathbbm{2}\mathbbm{1}}.
\end{split}
\end{align}
Using the objective variations~\eqref{delta_o}, the variation of the spatial Lagrange multiplier potential in~\eqref{spatial_LM_potential} reads:
\begin{align}
\begin{split}
\label{spatial_LM_potential_variation}
\delta \tilde{\pi}_{\vlambda} & =\delta \vlambda_{{\vr}}^T \hat{\vr}_{21} + \delta \vlambda_{{\vpsi}}^T \hat{\vpsi}_{21} 
\quad \quad \quad \quad \quad \quad \quad \quad \quad \quad \quad + \delta_o \vr_{{2}{1}}^T \vlambda_{{\vr}} \quad \quad \quad  \quad \quad \quad  \quad + \delta_o \vpsi_{21}^T \vlambda_{{\vpsi}} \\
& =\delta \vlambda_{{\vr}}^T \hat{\vr}_{21} + \delta \vlambda_{{\vpsi}}^T \hat{\vpsi}_{21} 
+ \underbrace{( \delta \vr_2^T \!-\! \delta \vr_1^T \!-\!  \delta \vtheta_1^T \! \vS[( \vr_2 - \vr_1 )] ) }_{\delta_o \vr_{{2}{1}}^T} \underbrace{\vlambda_{{\vr}}}_{\vf_2} + \underbrace{ ( \delta \vtheta_2^T - \delta \vtheta_1^T ) }_{\delta_o \vtheta_{21}^T} \underbrace{\vT^T(\vpsi_{21}) \vlambda_{{\vpsi}}}_{\vm_2}.
\end{split}
\end{align}
Similarly, using the variations~\eqref{delta_material}, the variation of the material Lagrange multiplier potential in~\eqref{spatial_LM_potential} reads:
\begin{align}
\begin{split}
\label{material_LM_potential_variation}
\delta \bar{\pi}_{\vlambda} & =\delta \vlambda_{{\vR}}^T \hat{\vR}_{\mathbbm{2}\mathbbm{1}} + \delta \vlambda_{{\vPsi}}^T \hat{\vPsi}_{\mathbbm{2}\mathbbm{1}} 
\quad \quad \quad \quad \quad \quad \quad \quad \quad \quad \quad \quad + \delta \vR_{\mathbbm{2}\mathbbm{1}}^T \vlambda_{{\vR}} \quad \,\,\,\, \quad \quad \quad  \quad \quad \quad  \quad + \delta \vPsi_{\mathbbm{2}\mathbbm{1}}^T \vlambda_{{\vPsi}} \\
& =\delta \vlambda_{{\vR}}^T \hat{\vR}_{\mathbbm{2}\mathbbm{1}} + \delta \vlambda_{{\vPsi}}^T \hat{\vPsi}_{\mathbbm{2}\mathbbm{1}} 
+ \underbrace{( \delta \vr_2^T \!-\! \delta \vr_1^T \!-\!  \delta \vtheta_1^T \! \vS[( \vr_2 - \vr_1 )] )\vLambda_1 }_{\delta \vR_{\mathbbm{2}\mathbbm{1}}^T} \underbrace{\vlambda_{{\vR}}}_{\vF_2} + \underbrace{ ( \delta \vtheta_2^T - \delta \vtheta_1^T ) \vLambda_1 }_{\delta \vtheta_{\mathbbm{2}\mathbbm{1}}^T} \underbrace{\vT^T(\vPsi_{\mathbbm{2}\mathbbm{1}}) \vlambda_{{\vPsi}}}_{\vM_2}.
\end{split}
\end{align}
In~\eqref{spatial_LM_potential_variation} and~\eqref{material_LM_potential_variation}, the terms on the left are the variational representation of the constraint equations~\eqref{spatial_penalty_potential0}, and the terms on the right represent the virtual work of the coupling forces and moments. With the relations $\vlambda_{{\vr}}=\vLambda_1 \vlambda_{{\vR}}$ and $\vlambda_{{\vpsi}}=\vLambda_1 \vlambda_{{\vPsi}}$ it can be concluded that these virtual work contributions are identical for the spatial and material representation. More specifically, the Lagrange multipliers $\vlambda_{{\vr}}$ and $\vlambda_{{\vR}}$ represent the spatial and material coupling forces. The coupling moments, however, are given by the Lagrange multipliers $\vlambda_{{\vpsi}}$ and $\vlambda_{{\vPsi}}$ multiplied with the corresponding transformation matrices $\vT^T(\vpsi_{21})$ and $\vT^T(\vPsi_{\mathbbm{2}\mathbbm{1}})$. In the recent contribution~\cite{steinbrecher2021}, it has been demonstrated how penalty and Lagrange multiplier potentials according to~\eqref{spatial_penalty_potential} and~\eqref{spatial_LM_potential} can be utilized for a consistent coupling of positions and rotations when embedding slender beams (1D Cosserat continua) into solid bodies (3D Boltzmann continua).


\section{Summary}
\label{sec:summary}

A universal framework has been proposed to formulate generalized section-section interaction potentials (SSIP) within the framework of the geometrically exact beam theory and to derive the resulting section-section interaction force and moment laws in a variationally consistent manner. While originally derived for the modeling of inter-molecular interactions (e.g., due to electrostatic, van der Waals or repulsive steric forces) between slender fibers, it has been demonstrated that these SSIPs can be identified as generalization of hyperelastic stored-energy functions underling the geometrically exact Simo-Reissner beam theory. Also, they allow to formulate general translational and rotational constraints when interpreting the associated Penalty or Lagrange multiplier potential as SSIP. In sum, the following main results have been derived:
\begin{itemize}
\item The interaction of two beams (lengths $l_1, l_2$, cross-sections $A_1,A_2$, molecule densities $\rho_1,\rho_2$) due to inter-molecular potentials $\Phi$ is described via a section-section interaction potential (SSIP) $\pi$:
\begin{align*}
  \Pi = \int_{l_1} \int_{l_2} \underbrace{\int_{A_1} \int_{A_2} \rho_1(\mb{\xi}_1) \rho_2(\mb{\xi}_2) \Phi(x) \dd A_2 \dd A_1}_{\pi} \dd s_2 \dd s_1.
\end{align*}

\item The SSIP $\tilde{\pi}(\vr_{21},\vpsi_{21})=\bar{\pi}(\vR_{\mathbbm{2}\mathbbm{1}},\vPsi_{\mathbbm{2}\mathbbm{1}})$ is defined by generalized spatial or material deformation measures describing the relative distance and rotation of two cross-sections (centroids $\vr_1, \vr_2$, triads $\vLambda_1, \vLambda_2$):
\begin{align*}
\begin{split}
  \vr_{21} &:= \vr_2 - \vr_1, \quad \vpsi_{21} := \text{rv}(\vLambda_2 \vLambda_1^\text{T}), \\
   \vR_{\mathbbm{2}\mathbbm{1}} &:= \vLambda_1^T \vr_{21}, \quad  \vPsi_{\mathbbm{2}\mathbbm{1}} := \vLambda_1^T \vpsi_{21}.
\end{split}
\end{align*}

\item The rotational deformation measures defined as multiplicative relative rotation (e.g., $\vpsi_{21} := \text{rv}(\vLambda_2 \vLambda_1^\text{T})$) satisfy objectivity, which is in contrast to obvious alternatives such as $\tilde{\vpsi}_{21}=\text{rv}(\vLambda_2)-\text{rv}(\vLambda_1)$.\\

\item The work-conjugated spatial and material section-section interaction forces and moments result as:
\begin{align*}
\begin{split}
\vf_2&=\left(\frac{\partial \tilde{\pi}}{\partial \vr_{21}}\right)^{\!\!T}, \quad \quad \vm_2= \left(\frac{\partial_m \tilde{\pi}}{\partial_m \vtheta_{21}}\right)^{\!\!T} :=\vT^T\!(\vpsi_{21})\left(\frac{\partial \tilde{\pi}}{\partial \vpsi_{21}}\right)^{\!\!T},\\
\vF_2&=\left(\frac{\partial \bar{\pi}}{\partial \vR_{\mathbbm{2}\mathbbm{1}}}\right)^{\!\!T}, \quad \quad \vM_2=\left(\frac{\partial_m \bar{\pi}}{\partial_m \vtheta_{\mathbbm{2}\mathbbm{1}}}\right)^{\!\!T}:=\vT^T\!(\vPsi_{\mathbbm{2}\mathbbm{1}})\left(\frac{\partial \bar{\pi}}{\partial \vPsi_{\mathbbm{2}\mathbbm{1}}}\right)^{\!\!T}.
\end{split}
\end{align*}
Importantly, the interaction moments are defined as \textit{multiplicative derivative} $\partial_m (.) / \partial_m \vtheta$ of the SSIP.

\item Hyperelastic stored-energy functions relating deformation measures and stress-resultants of the geometrically exact Simo-Reissner (SR) beam theory can be identified as asymptotic limit of the proposed SSIPs for infinitesimally small relative distances and rotations between the interacting cross-sections:
\begin{align*}
\Pi_{int, SR}= \int_l \bar{\pi}_{int,SR}(\mb{\Gamma},\mb{\Omega}) d s = 
\lim_{\Delta \rightarrow 0} \int \limits_{0}^{l_1} \int \limits_{s_1- \Delta}^{s_1+\Delta} \bar{\pi}(\hat{\vR}_{\mathbbm{2}\mathbbm{1}},\hat{\vPsi}_{\mathbbm{2}\mathbbm{1}}) \dd s_2 \dd s_1.
\end{align*}

\item General translational and rotational constraints of the form $\hat{\vr}_{21}:=\vr_{21}-\vr_{21}^0 \dot{=}0$ and $\hat{\vpsi}_{21}=\vpsi_{21}-\vpsi_{21}^0 \dot{=}0$
can be enforced by penalty or Lagrange multiplier potentials representing a special case of SSIP laws:
\begin{align*}
\begin{split}
\tilde{\pi}_{\epsilon}(\hat{\vr}_{21},\hat{\vpsi}_{21})& =\frac{1}{2}\epsilon_{\vr} \hat{\vr}_{21}^T \hat{\vr}_{21} + \frac{1}{2}\epsilon_{\vpsi} \hat{\vpsi}_{21}^T \hat{\vpsi}_{21} \\
\tilde{\pi}_{\vlambda}(\hat{\vr}_{21},\hat{\vpsi}_{21})& =\vlambda_{{\vr}}^T \hat{\vr}_{21} + \vlambda_{{\vpsi}}^T \hat{\vpsi}_{21}.
\end{split}
\end{align*}
\end{itemize}
Future applications of the proposed universal framework include fiber-based structures and materials in technical and biological systems, where it allows to model short- or long-ranged inter-molecular (e.g., electrostatic, van der Waals or repulsive steric) interactions between fibers in geometrically complex arrangements and to formulate 
translational and rotational coupling constraints between different fibers (e.g., cross-linked polymer chains) or between fibers and a matrix phase (e.g., fiber-reinforced composites).

\appendix

\section{Useful identities for large rotations}
\label{appendix:utilities}

The following utility equations have been employed in the derivations shown in the main text:
\begin{align}
\label{A_Lambda1}
  \vLambda(\vLambda_1 \vpsi) = \vLambda_1  \vLambda(\vpsi) \vLambda_1^T \,\,\, \forall \,\,\, { \boldsymbol{\psi} } \in \Re^3, \, \vLambda, \vLambda_1  \in S \! O(3)
\end{align}
\begin{align}
\label{A_T1}
  \vT(\vLambda_1 \vpsi) = \vLambda_1  \vT(\vpsi) \vLambda_1^T \quad \text{or} \quad \vT^T(\vLambda_1 \vpsi) = \vLambda_1  \vT^T(\vpsi) \vLambda_1^T
\end{align}
\begin{align}
\label{A_T2}
  \vT^T(\vtheta) = \vT(\vtheta) \vLambda (\vtheta)
\end{align}
\begin{align}
\label{A_T3}
  \vT\!(\vpsi) \vpsi = \vT^T\!(\vpsi) \vpsi = \vpsi
\end{align}

\section{Variation of the spatial and material deformation measures}
\label{appendix:variations}

In the following, a brief derivation shall be presented for the variations~$\delta \vR_{\mathbbm{2}\mathbbm{1}}, \delta \vPsi_{\mathbbm{2}\mathbbm{1}}$ occurring in~\eqref{delta_material}.
First, starting from the definition~\eqref{D_and_Psi}, we immediately arrive at an expression for the variation of the material relative position vector $\vR_{\mathbbm{2}\mathbbm{1}}$ by employing the product rule as well as~\eqref{largerotations_deltalambdaspatial}:
\begin{align}
  \delta \vR_{\mathbbm{2}\mathbbm{1}} &= - \vLambda_1^T \, \vS(\delta \vtheta_1) ( \vr_2 - \vr_1 ) + \vLambda_1^T (\delta \vr_2 - \delta \vr_1) \\
             &= \vLambda_1^T  (\delta \vr_2- \delta \vr_1 - \delta \vtheta_1 \times (\vr_2-\vr_1)   ).
\end{align}

The derivation is a bit more involved for the material relative rotation vector $\delta \vPsi_{\mathbbm{2}\mathbbm{1}}$. We aim to calculate the total variation of the material vector $\vPsi_{\mathbbm{2}\mathbbm{1}}$ in the sense of an additive change of this vector as consequence of multiplicative variations $\delta \vtheta_1$ and $\delta \vtheta_2$ of the orientations of $\vLambda_1$ and $\vLambda_2$. Since $\vPsi_{\mathbbm{2}\mathbbm{1}}$ represents a rotation angle with associated triad $\vLambda_{\mathbbm{2}\mathbbm{1}}(\vPsi_{\mathbbm{2}\mathbbm{1}})$, the sought-after total variation of $\vPsi_{\mathbbm{2}\mathbbm{1}}$ can be interpreted as an additive variation. This additive variation can alternatively be expressed via a material/right multiplicative variation $\delta \vTheta_{\mathbbm{2}\mathbbm{1}}$ associated with $\delta \vLambda_{\mathbbm{2}\mathbbm{1}} = \vLambda_{\mathbbm{2}\mathbbm{1}} \vS(\delta \vTheta_{\mathbbm{2}\mathbbm{1}})$. According to~\eqref{largerotations_tmatrix2}, the relation between an additive and material/right multiplicative variation is given as:
\begin{align}
\label{var_def_material_triad0}
  \delta \vPsi_{\mathbbm{2}\mathbbm{1}}=\mb{T}^T(\vPsi_{\mathbbm{2}\mathbbm{1}}) \delta \vTheta_{\mathbbm{2}\mathbbm{1}}.
\end{align}
This relation will be required later. Next, variation of both sides of the defining equation~\eqref{Lambda_Psi} yields
\begin{align}
\label{var_def_material_triad}
  \vLambda_{\mathbbm{2}\mathbbm{1}} \vS(\delta \vTheta_{\mathbbm{2}\mathbbm{1}}) = - \vS \left( \delta \vTheta_{1}  \right) \underbrace{\vLambda_1^T \vLambda_2}_{\vLambda_{\mathbbm{2}\mathbbm{1}}} +  \underbrace{\vLambda_1^T \vLambda_2}_{\vLambda_{\mathbbm{2}\mathbbm{1}}} \vS \left( \delta \vTheta_{2}  \right),
\end{align}
where $\delta \vTheta_{\mathbbm{2}\mathbbm{1}}$, $\delta \vTheta_{1}$ and $\delta \vTheta_{2}$ are the right/material multiplicative variations associated with the triads $\vLambda_{\mathbbm{2}\mathbbm{1}}$, $\vLambda_1$ and $\vLambda_2$. Left-multiplication of~\eqref{var_def_material_triad} with $\vLambda_{\mathbbm{2}\mathbbm{1}}^T$ together with the definition $\vLambda_{\mathbbm{2}\mathbbm{1}}=\vLambda_1^T \vLambda_2$ yields:
\begin{align}
\label{var_def_material_trias2}
 \vS(\delta \vTheta_{\mathbbm{2}\mathbbm{1}}) = - \vLambda_{\mathbbm{2}\mathbbm{1}}^T \vS \left( \delta \vTheta_{1}  \right) \vLambda_{\mathbbm{2}\mathbbm{1}}  +  \vS \left( \delta \vTheta_{2}  \right),
\end{align}
Eventually, by using $\vLambda_{\mathbbm{2}\mathbbm{1}}^T \vS \left( \delta \vTheta_{1}  \right) \vLambda_{\mathbbm{2}\mathbbm{1}} = \vS \left( \vLambda_{\mathbbm{2}\mathbbm{1}}^T  \delta \vTheta_{1}  \right)$ the following relation can be derived from~\eqref{var_def_material_trias2}:
\begin{align}
\label{var_def_material_trias3}
 \delta \vTheta_{\mathbbm{2}\mathbbm{1}} = \delta \vTheta_{2} - \vLambda_{\mathbbm{2}\mathbbm{1}}^T  \delta \vTheta_{1}.
\end{align}
Substituting this result into~\eqref{var_def_material_triad0} by using~\eqref{A_T2}, $\vLambda_{\mathbbm{2}\mathbbm{1}}=\vLambda_1^T \vLambda_2$ as well as $\delta \vtheta_{i}=\vLambda_i \delta \vTheta_{i}$ ($i=1,2$) yields:
\begin{align}
  \delta \vPsi_{\mathbbm{2}\mathbbm{1}} &= \vdT^T(\vPsi_{\mathbbm{2}\mathbbm{1}}) \left( \delta \vTheta_{2} - \vLambda_{\mathbbm{2}\mathbbm{1}}^T  \delta \vTheta_{1} \right) \\
  &= \vdT(\vPsi_{\mathbbm{2}\mathbbm{1}}) \left( \vLambda_{\mathbbm{2}\mathbbm{1}} \delta \vTheta_{2} -  \delta \vTheta_{1} \right) \\
  &= \vdT(\vPsi_{\mathbbm{2}\mathbbm{1}}) \vLambda_1^T \left(  \delta \vtheta_{2} - \delta \vtheta_1 \right).
\end{align}

\bibliography{library}  

\begin{thebibliography}{10}
\expandafter\ifx\csname url\endcsname\relax
  \def\url#1{\texttt{#1}}\fi
\expandafter\ifx\csname urlprefix\endcsname\relax\def\urlprefix{URL }\fi
\expandafter\ifx\csname href\endcsname\relax
  \def\href#1#2{#2} \def\path#1{#1}\fi

\bibitem{durville2010}
D.~Durville, {Simulation of the mechanical behaviour of woven fabrics at the
  scale of fibers}, International Journal of Material Forming 3~(2) (2010)
  1241--1251.

\bibitem{Kulachenko2012}
A.~Kulachenko, T.~Uesaka, {Direct simulations of fiber network deformation and
  failure}, Mechanics of Materials 51 (2012) 1--14.

\bibitem{Weeger2016}
O.~Weeger, Y.~S.~B. Kang, S.-K. Yeung, M.~L. Dunn, {Optimal Design and
  Manufacture of Active Rod Structures with Spatially Variable Materials}, 3D
  Printing and Additive Manufacturing 3~(4) (2016) 204--215.

\bibitem{Meier2017b}
C.~Meier, M.~J. Grill, W.~A. Wall, A.~Popp, {Geometrically exact beam elements
  and smooth contact schemes for the modeling of fiber-based materials and
  structures}, International Journal of Solids and Structures 154 (2018)
  124--146.

\bibitem{Pattinson2019}
S.~W. Pattinson, M.~E. Huber, S.~Kim, J.~Lee, S.~Grunsfeld, R.~Roberts,
  G.~Dreifus, C.~Meier, L.~Liu, N.~Hogan, et~al., Additive manufacturing of
  biomechanically tailored meshes for compliant wearable and implantable
  devices, Advanced Functional Materials 29~(32) (2019) 1901815.

\bibitem{Mattheij2000}
P.~Mattheij, K.~Gliesche, D.~Feltin, {3D} reinforced stitched carbon/epoxy
  laminates made by tailored fibre placement, Composites Part A: Applied
  Science and Manufacturing 31~(6) (2000) 571--581.

\bibitem{durville2007}
D.~Durville, {Finite Element Simulation of Textile Materials at Mesoscopic
  Scale}, in: Finite element modelling of textiles and textile composites,
  Saint-Petersbourg, Russian Federation, 2007, pp. 1--14.

\bibitem{steinbrecher2020}
I.~Steinbrecher, M.~Mayr, M.~J. Grill, J.~Kremheller, C.~Meier, A.~Popp, A
  mortar-type finite element approach for embedding 1d beams into 3d solid
  volumes, Computational Mechanics 66~(6) (2020) 1377--1398.

\bibitem{Khristenko2021}
U.~Khristenko, S.~Schu{\ss}, M.~Kr{\"u}ger, F.~Schmidt, B.~Wohlmuth, C.~Hesch,
  Multidimensional coupling: A variationally consistent approach to
  fiber-reinforced materials, Computer Methods in Applied Mechanics and
  Engineering 382 (2021) 113869.

\bibitem{Castro2011}
C.~E. Castro, F.~Kilchherr, D.-N. Kim, E.~L. Shiao, T.~Wauer, P.~Wortmann,
  M.~Bathe, H.~Dietz, {A primer to scaffolded DNA origami}, Nature Methods
  8~(3) (2011) 221--229.

\bibitem{Gautieri2012}
A.~Gautieri, M.~I. Pate, S.~Vesentini, A.~Redaelli, M.~J. Buehler, {Hydration
  and distance dependence of intermolecular shearing between collagen molecules
  in a model microfibril}, Journal of Biomechanics 45~(12) (2012) 2079--2083.

\bibitem{Sauer2009}
R.~A. Sauer, {Multiscale modelling and simulation of the deformation and
  adhesion of a single gecko seta}, Computer Methods in Biomechanics and
  Biomedical Engineering 12~(6) (2009) 627--640.

\bibitem{lindstrom2010biopolymer}
S.~B. Lindstr{\"o}m, D.~A. Vader, A.~Kulachenko, D.~A. Weitz, Biopolymer
  network geometries: Characterization, regeneration, and elastic properties,
  Physical Review E 82~(5) (2010) 051905.

\bibitem{muller2015resolution}
K.~W. M{\"u}ller, C.~Meier, W.~A. Wall, Resolution of sub-element length scales
  in brownian dynamics simulations of biopolymer networks with geometrically
  exact beam finite elements, Journal of Computational Physics 303 (2015)
  185--202.

\bibitem{Negi2018}
V.~Negi, R.~C. Picu, {Mechanical behavior of cross-linked random fiber networks
  with inter-fiber adhesion}, Journal of the Mechanics and Physics of Solids
  122 (2018) 418--434.

\bibitem{Goodrich2018}
C.~P. Goodrich, M.~P. Brenner, K.~Ribbeck, {Enhanced diffusion by binding to
  the crosslinks of a polymer gel}, Nature Communications 9~(1) (2018) 4348.

\bibitem{GrillParticleMobilityHydrogels}
M.~J. Grill, J.~F. Eichinger, J.~Koban, C.~Meier, O.~Lieleg, W.~A. Wall, {A
  novel modelling and simulation approach for the hindered mobility of charged
  particles in biological hydrogels}, Proceedings of the Royal Society A:
  Mathematical, Physical and Engineering Sciences 477~(2249) (2021) 20210039.

\bibitem{GrillPeelingPulloff}
M.~J. Grill, C.~Meier, W.~A. Wall, Investigation of the peeling and pull-off
  behavior of adhesive elastic fibers via a novel computational beam
  interaction model, The Journal of Adhesion 97~(8) (2021) 730--759.

\bibitem{Eichinger2021}
J.~F. Eichinger, M.~J. Grill, I.~D. Kermani, R.~C. Aydin, W.~A. Wall, J.~D.
  Humphrey, C.~J. Cyron, A computational framework for modeling cell--matrix
  interactions in soft biological tissues, Biomechanics and modeling in
  mechanobiology 20~(5) (2021) 1851--1870.

\bibitem{BundlesPNAS}
V.~M. Slepukhin, M.~J. Grill, Q.~Hu, E.~L. Botvinick, W.~A. Wall, A.~J. Levine,
  {Topological defects produce kinks in biopolymer filament bundles},
  Proceedings of the National Academy of Sciences 118~(15) (2021) e2024362118.

\bibitem{Argento1997}
C.~Argento, A.~Jagota, W.~C. Carter, {Surface formulation for molecular
  interactions of macroscopic bodies}, Journal of the Mechanics and Physics of
  Solids 45~(7) (1997) 1161--1183.

\bibitem{Sauer2007a}
R.~A. Sauer, S.~Li, {A contact mechanics model for quasi-continua},
  International Journal for Numerical Methods in Engineering 71~(8) (2007)
  931--962.

\bibitem{Sauer2009a}
R.~A. Sauer, P.~Wriggers, {Formulation and analysis of a three-dimensional
  finite element implementation for adhesive contact at the nanoscale},
  Computer Methods in Applied Mechanics and Engineering 198~(49) (2009)
  3871--3883.

\bibitem{Sauer2013}
R.~A. Sauer, L.~{De Lorenzis}, {A computational contact formulation based on
  surface potentials}, Computer Methods in Applied Mechanics and Engineering
  253 (2013) 369--395.

\bibitem{Fan2015}
H.~Fan, S.~Li, {A three-dimensional surface stress tensor formulation for
  simulation of adhesive contact in finite deformation}, International Journal
  for Numerical Methods in Engineering 107~(3) (2016) 252--270.

\bibitem{Du2019}
S.~Du, H.~{Ben Dhia}, {An asymptotic numerical method to solve compliant
  Lennard-Jones-based contact problems involving adhesive instabilities},
  Computational Mechanics 63~(6) (2019) 1261--1281.

\bibitem{Mergel2019}
J.~C. Mergel, R.~Sahli, J.~Scheibert, R.~A. Sauer, {Continuum contact models
  for coupled adhesion and friction}, The Journal of Adhesion 95~(12) (2019)
  1101--1133.

\bibitem{wriggers1997}
P.~Wriggers, G.~Zavarise, {On contact between three-dimensional beams
  undergoing large deflections}, Communications in Numerical Methods in
  Engineering 13~(6) (1997) 429--438.

\bibitem{litewka2005}
P.~Litewka, {The penalty and Lagrange multiplier methods in the frictional 3d
  beam-to-beam contact problem}, Civil and Environmental Engineering Reports 1
  (2005) 189--207.

\bibitem{Chamekh2014}
M.~Chamekh, S.~Mani-Aouadi, M.~Moakher, {Stability of elastic rods with
  self-contact}, Computer Methods in Applied Mechanics and Engineering 279
  (2014) 227--246.

\bibitem{GayNeto2016a}
A.~{Gay Neto}, P.~M. Pimenta, P.~Wriggers, {A master-surface to master-surface
  formulation for beam to beam contact. Part I: Frictionless interaction},
  Computer Methods in Applied Mechanics and Engineering 303 (2016) 400--429.

\bibitem{Konyukhov2016}
A.~Konyukhov, O.~Mrenes, K.~Schweizerhof, {Consistent Development of a
  Beam-To-Beam Contact Algorithm via the Curve-to-Solid Beam Contact ?-
  Analysis for the Nonfrictional Case}, International Journal for Numerical
  Methods in Engineering 113~(7) (2018) 1108--1144.

\bibitem{Weeger2017}
O.~Weeger, B.~Narayanan, L.~{De Lorenzis}, J.~Kiendl, M.~L. Dunn, {An
  isogeometric collocation method for frictionless contact of Cosserat rods},
  Computer Methods in Applied Mechanics and Engineering 321 (2017) 361--382.

\bibitem{meier2016}
C.~Meier, A.~Popp, W.~A. Wall, {A finite element approach for the line-to-line
  contact interaction of thin beams with arbitrary orientation}, Computer
  Methods in Applied Mechanics and Engineering 308 (2016) 377--413.

\bibitem{Meier2017}
C.~Meier, W.~A. Wall, A.~Popp, A unified approach for beam-to-beam contact,
  Computer Methods in Applied Mechanics and Engineering 315 (2017) 972--1010.

\bibitem{bosten2022mortar}
A.~Bosten, A.~Cosimo, J.~Linn, O.~Br{\"u}ls, A mortar formulation for
  frictionless line-to-line beam contact, Multibody System Dynamics 54~(1)
  (2022) 31--52.

\bibitem{Sauer2014}
R.~A. Sauer, J.~C. Mergel, {A geometrically exact finite beam element
  formulation for thin film adhesion and debonding}, Finite Elements in
  Analysis and Design 86 (2014) 120--135.

\bibitem{Schmidt2015}
M.~G. Schmidt, A.~E. Ismail, R.~A. Sauer, {A continuum mechanical surrogate
  model for atomic beam structures}, International Journal for Multiscale
  Computational Engineering 13~(5) (2015) 413--442.

\bibitem{GrillSSIP}
M.~J. Grill, W.~A. Wall, C.~Meier, {A computational model for molecular
  interactions between curved slender fibers undergoing large 3D deformations
  with a focus on electrostatic, van der Waals, and repulsive steric forces},
  International Journal for Numerical Methods in Engineering 121~(10) (2020)
  2285--2330.

\bibitem{Grill2022a}
M.~J. Grill, W.~A. Wall, C.~Meier, {Asymptotically consistent and
  computationally efficient modeling of short-ranged molecular interactions
  between curved slender fibers undergoing large 3D deformations}, submitted
  for publication, arXiv preprint arXiv:2208.03149.

\bibitem{Grill2022b}
M.~J. Grill, W.~A. Wall, C.~Meier, {Analytical disk-cylinder interaction
  potential laws for the computational modeling of adhesive, deformable
  (nano)fibers}, submitted for publication, arXiv preprint arXiv:2208.03074.

\bibitem{reissner1972}
E.~Reissner, {On one-dimensional finite-strain beam theory: The plane problem},
  Zeitschrift f{\"{u}}r Angewandte Mathematik und Physik (ZAMP) 23~(5) (1972)
  795--804.

\bibitem{simo1985}
J.~C. Simo, {A finite strain beam formulation. The three-dimensional dynamic
  problem. Part I}, Computer Methods in Applied Mechanics and Engineering 49
  (1985) 55--70.

\bibitem{simo1986}
J.~C. Simo, L.~Vu-Quoc, {A three-dimensional finite strain rod model. Part II:
  Computational aspects}, Computer Methods in Applied Mechanics and Engineering
  58 (1986) 79--116.

\bibitem{Cardona1988}
A.~Cardona, M.~Geradin, {A beam finite element non-linear theory with finite
  rotations}, International Journal for Numerical Methods in Engineering
  26~(11) (1988) 2403--2438.

\bibitem{ibrahimbegovic1995computational}
A.~Ibrahimbegovi{\'c}, F.~Frey, I.~Ko{\v{z}}ar, Computational aspects of
  vector-like parametrization of three-dimensional finite rotations,
  International Journal for Numerical Methods in Engineering 38~(21) (1995)
  3653--3673.

\bibitem{Crisfield1999}
M.~A. Crisfield, G.~Jeleni{\'{c}}, {Objectivity of strain measures in the
  geometrically exact three-dimensional beam theory and its finite-element
  implementation}, Proceedings of the Royal Society of London. Series A:
  Mathematical, Physical and Engineering Sciences 455 (1999) 1125--1147.

\bibitem{jelenic1999}
G.~Jeleni\'{c}, M.~A. Crisfield, {Geometrically exact 3D beam theory:
  Implementation of a strain-invariant finite element for statics and
  dynamics}, Computer Methods in Applied Mechanics and Engineering 171~(1--2)
  (1999) 141--171.

\bibitem{betsch2002frame}
P.~Betsch, P.~Steinmann, Frame-indifferent beam finite elements based upon the
  geometrically exact beam theory, International journal for numerical methods
  in engineering 54~(12) (2002) 1775--1788.

\bibitem{leyendecker2006objective}
S.~Leyendecker, P.~Betsch, P.~Steinmann, Objective energy--momentum conserving
  integration for the constrained dynamics of geometrically exact beams,
  Computer Methods in Applied Mechanics and Engineering 195~(19-22) (2006)
  2313--2333.

\bibitem{romero2004}
I.~Romero, The interpolation of rotations and its application to finite element
  models of geometrically exact rods, Computational Mechanics 34~(2) (2004)
  121--133.

\bibitem{romero2008}
I.~Romero, {A comparison of finite elements for nonlinear beams: the absolute
  nodal coordinate and geometrically exact formulations}, Multibody System
  Dynamics 20~(1) (2008) 51--68.

\bibitem{cesarek2012}
P.~\v{C}e\v{s}arek, M.~Saje, D.~Zupan, {Kinematically exact curved and twisted
  strain-based beam}, International Journal of Solids and Structures 49~(13)
  (2012) 1802--1817.

\bibitem{Bauchau2014}
O.~A. Bauchau, S.~Han, A.~Mikkola, M.~K. Matikainen, {Comparison of the
  absolute nodal coordinate and geometrically exact formulations for beams},
  Multibody System Dynamics 32~(1) (2014) 67--85.

\bibitem{sonneville2014}
V.~Sonneville, A.~Cardona, O.~Br\"uls, {Geometrically exact beam finite element
  formulated on the special Euclidean group}, Computer Methods in Applied
  Mechanics and Engineering 268 (2014) 451--474.

\bibitem{Meier2014}
C.~Meier, A.~Popp, W.~A. Wall, {An objective 3D large deformation finite
  element formulation for geometrically exact curved Kirchhoff rods}, Computer
  Methods in Applied Mechanics and Engineering 278 (2014) 445--478.

\bibitem{Meier2019}
C.~Meier, A.~Popp, W.~A. Wall, Geometrically {E}xact {F}inite {E}lement
  {F}ormulations for {S}lender {B}eams: {K}irchhoff--{L}ove {T}heory {V}ersus
  {S}imo--{R}eissner {T}heory, Archives of Computational Methods in Engineering
  26~(1) (2019) 163--243.

\bibitem{spurrier1978}
R.~A. Spurrier, {Comment on ''singularity-free extraction of a quaternion from
  a direction-cosine matrix''}, Journal of Spacecraft and Rockets 15 (1978)
  255--255.

\bibitem{antmann1995}
S.~S. Antmann, {Nonlinear Problems of Elasticity}, Springer, 1995.

\bibitem{steinbrecher2021}
I.~Steinbrecher, A.~Popp, C.~Meier, Consistent coupling of positions and
  rotations for embedding 1d cosserat beams into 3d solid volumes,
  Computational Mechanics (2021) 1--32.

\end{thebibliography}

\end{document}